\documentstyle[10pt,aaspp4] {article}
\def\deg{$^{\circ}\,$}

\def\etal{{\it et~al.\ }}

\begin{document} 

\title{The Asymmetry of Galaxies: Physical Morphology for Nearby and
High Redshift Galaxies}

\author{Christopher J. Conselice, Matthew A. Bershady}

\affil{Department of Astronomy, University of Wisconsin, Madison, 475
N. Charter St. Madison, WI, 53706-1582 (chris@astro.wisc.edu,
mab@astro.wisc.edu)}

\author{Anna Jangren}
 
\affil{Department of Astronomy and Astrophysics, Pennsylvania State
University, 525 Davey Lab, University Park, PA, 16802
(jangren@astro.psu.edu)}


\begin{abstract}

We present a detailed study of rotational asymmetry in galaxies for
both morphological and physical diagnostic purposes. An unambiguous
method for computing asymmetry is developed, robust for both distant
and nearby galaxies. By degrading real galaxy images, we test the
reliability of this asymmetry measure over a range of observational
conditions, e.g. spatial resolution and signal-to-noise
(S/N). Compared to previous methods, this new algorithm avoids the
ambiguity associated with choosing a center by using a minimization
method, and successfully corrects for variations in S/N. There is,
however, a strong relationship between the rotational asymmetry and
physical resolution (distance at fixed spatial resolution); objects
become more symmetric when less well-resolved.

We further investigate asymmetry as a function of galactic radius and
rotation. We find the asymmetry index has a strong radial dependence
that differs vastly between Hubble types. As a result, {\it a
meaningful asymmetry index must be specified within a well-defined
radius representative of the physical galaxy scale.} We enumerate
several viable alternatives, which excludes the use of
isophotes. Asymmetry as a function of angle (A$_\phi$) is also a
useful indicator of ellipticity and higher-order azimuthal structure.
In general, we show the power of asymmetry as a morphological
parameter lies in the strong correlation with $(B-V)$ color for
galaxies undergoing normal star formation, spanning all Hubble types
from ellipticals to irregular galaxies. Interacting galaxies do not
fall on this asymmetry-color ``fiducial sequence,'' as these galaxies
are too asymmetric for their color. We propose to use this fact to
distinguish between `normal' galaxies and galaxies undergoing an
interaction or merger at high redshift.
  
\end{abstract}

\newpage

\section{Introduction}

\subsection{Galaxy Morphology}

Ever since galaxies were recognized as distinct physical systems, one
of the main goals in extragalactic astronomy has been to characterize
their forms, or morphology, and to determine how this classification
relates to physical properties. This basic taxonomical process is
indeed the basis for any observational science.  The first attempts at
classification were on a subjective level, and began with the work of
Curtis (1918), Hubble (1926, 1936), and Sandage (1961). As more images
of galaxies became available, the morphological system developed by
Hubble was generally adopted by all astronomers, and later refined by
van den Bergh (1960a, 1960b) and de Vaucoulers (1959). Other
morphological systems were also developed by Morgan (1958) based on
the correlation of the physical characteristics of galaxy spectra with
the concentration of the light profiles. Yet since the time of
Hubble's 1926 work, the system of morphology for galaxies has changed
little. When Hubble developed his original morphological system, his
sample consisted of mostly nearby, luminous galaxies, with only 3\%
``irregular'' galaxies. These galaxies were not well incorporated in
his sequence, but for his purposes, the morphological system developed
was adequate for classifying 97\% of his sample.

As deeper galaxy catalogues emerged, however, more and more
galaxies fell into the catch-all, ``irregular'' morphological class.
Today, Hubble Space Telescope (HST) imaging reveals that a large
fraction of distant galaxies have morphologies that do not fit into
the the elliptical - spiral Hubble sequence.  The Hubble sequence also
fails to be useful when classifying galaxies in clusters, with most
galaxies classified as S0 or E -- classifications which fail to
account of the wide range of cluster galaxy properties (e.g. Koopmann
\& Kinney 1998). Spectral parameters are often more useful in these
cases (e.g. Dressler \& Gunn 1992). Today we classify irregular
galaxies not simply in a morphological system, but with regard to the
physical mechanisms in operation or the salient physical conditions
(e.g. starburst galaxies, interacting galaxies, gas-rich and
gas-poor). A morphological classification which reflects these
physical differences would be a powerful tool for studying the
mechanisms driving galaxy evolution. Such studies naturally must
include high redshift galaxies. Therefore, a morphological system that
encompasses all galaxies, and works sensibly over a wide range in
redshift is absolutely essential, but at present does not exist.

Recently, new methods of classifying galaxies have been proposed. One
line of effort has been to train artificial neural networks to
reproduce the Hubble scheme in an objective way (Burda \& Feitzinger
1992, Storrie-Lombardi \etal 1992, Serra-Ricart \etal 1993, Odewahn
1995, Naim \etal 1995, Odewahn \etal 1996). Spiekermann's (1992)
approach using fuzzy logic was along this line.  The number of galaxy
images in modern surveys, such as the ``Sloan Digital Sky Survey''
will be enormous, and hence such automatable methods of morphological
classification are desirable.  However, a Hubble classification
carries with it the limitations mentioned above; a system that can
classify galaxies in a straightforward and quantitative manner that is
based on a sound physical and morphological basis would be
preferable.

Another approach to galaxy classification has been to develop sets of
quantitative measures of the bulk image structures of galaxies. These
methods have the potential to either replace, modify, or improve the
current Hubble scheme. The new classifications generally rely on a set
of photometric and/or spectral properties that are internally
correlated, and correspond also with the apparent morphology of
galaxies.  Morgan's (1958, 1959) use of central light concentration
was the first example of such a classification. Indeed, the use of
concentration indices for inferring the morphology of galaxies has
continued to improve, and along with surface-brightness and asymmetry
has become one of the major tools for classifying both nearby and
distant galaxies (e.g. Okamura, Kodaira, \& Watanabe 1984; Doi \etal
1993; Abraham \etal 1994; Jangren \etal 1999). A different method --
applicable for spirals -- has been suggested by Elmegreen \& Elmegreen
(1982): measures of spiral arm morphology, particularly their
patchiness, can be used for classification. Related attempts to
classify galaxies have included the use of principle component
analysis of photometric structures (Whitmore 1984; Watanabe \etal
1985; Han 1995). These systems revealed correlations of physical and
morphological features of galaxies, but have not been generally
adopted for practical use, and the basic Hubble (1926) system still
lives on.

A key element missing from recent work listed above is the connection
made by Morgan between image structure and stellar content
(i.e. between light concentration, or central surface-brightness, and
spectral type). Ironically, in parallel to the above efforts to
quantify image structure, there has been considerable effort to
develop quantitative methods of spectral classification based on
broad-band colors (Bershady 1995) and spectra (Connolly \etal 1995,
Zaritsky \etal 1995, Folkes \etal 1996, Bromley \etal 1998, Ronen
\etal 1999). What is needed, then, is to go full circle to where
Morgan left off, by tying together the spectral types with the
quantitative classification based image structure.  Here, we propose
that using a measure of asymmetry and color for galaxies is a powerful
method towards accomplishing this goal.  In an accompanying paper
(Jangren \etal 1999) we explore the additional parameters of size,
surface-brightness and image concentration.

\subsection{Asymmetry as a Physical Classifier}

Symmetry has always been one the most basic features and assumptions
of most galaxy morphology systems, but also one of the most overlooked
for more detailed study.  The earliest galaxy morphology papers by
Curtis (1918) and Hubble (1926) described galaxies in terms of their
symmetry, in most cases a 180\deg symmetry. Hubble (1926) describes
elliptical and spirals galaxies in their most basic terms as systems
``characterized by rotational symmetry about dominating nuclei''. In
fact, it is striking how symmetric galaxy systems are, and as we will
show, almost always have a minimum asymmetry at a 180\deg rotation
angle. 

Models of galaxies often assume that the mass distribution of a galaxy
is symmetric. Galaxies are, to first order, dynamically relaxed
systems. Understanding how and in what manner the distribution of
galaxy light is asymmetric can help reveal dynamical processes in
galaxies.  For example, galaxies disturbed by interactions or mergers
will tend to have large asymmetries. For quite some time there has
been considerable effort to characterize the asymmetry in HI gas in
spiral galaxies (e.g. Baldwin \etal 1980; Richter \& Sancisi 1994);
attempts to do this in optical light are relatively recent (e.g. Rix
\& Zaritsky 1995; Kornreich, Haynes, \& Lovelace 1998; Rudnick \& Rix
1998). While HI-studies benefit directly from kinematic data, optical
studies offer complementary information since optical photons
predominantly come from stars, which are collisionless, and are
believed to trace well the underlying matter distribution in the disk.

The quantitative use of asymmetry as a morphological parameter was used
first by Schade \etal (1995) as a characterization of distant galaxies
observed with the Hubble Space Telescope (HST). Further use of
symmetry for galaxies in deep HST images has been carried out by
Abraham \etal (1996a, 1996b) and van den Bergh \etal (1996).  These
papers, however, use asymmetry only as a crude, type-characterization
of distant galaxies in the framework of the Hubble Sequence.

Attempts to characterize asymmetry for nearby galaxies, and its
usefulness as a morphological parameter within existing frameworks was
first carried out by Conselice (1997; hereafter C97). In C97 it was
shown that asymmetry increased with Hubble Type, but with a large
spread. Potentially more important was the strong correlation found
between color and asymmetry, and a lack of a strong correlation
between luminosity and asymmetry for the narrow absolute magnitude of
the C97 sample. 

In this paper we investigate further the relationship between
asymmetry and other physical parameters, such as color and luminosity,
and the usefulness of asymmetry as a morphological parameter. From
these results, our expectations are that asymmetry can be incorporated
as a pillar of a new classification system which better describes and
correlates with physical features and parameters than the Hubble
sequence.

The paper proceeds as follows. The calibration data set of local
galaxies used for this study is described in the next section.  In
\S3, we then compare different methods for computing asymmetry, and
propose a new procedure for measuring asymmetry as a robust quantity.
This method includes an iterative scheme for finding the center of
rotation (\S3.3), a noise correction (\S3.4), and a well-defined
radius of extraction.  We demonstrate how asymmetry changes as a
function of both the rotation angle used to compute it (\S4.1), and as
a function of galactic radius (\S4.2). We also discuss the
correlation between asymmetry and other physical parameters, i.e.
the $(B-V)$ color of a galaxy (\S4.3), the Hubble type, and the
concentration of light (\S4.4). Resolution effects are considered in
\S5.1. In \S4.3 we also discuss the causes of asymmetry, concluding
that asymmetries are produced by either star formation in spiral arms,
or dynamical effects related to interactions with other galaxies. We
show these two causes can be distinguished by examining their position
in a color-asymmetry diagram. These results are directly applicable to
distant galaxies with resolved structure, such as those in the Hubble
Deep Field (HDF).

\section{Calibration Data Set}

The data used in this present study are the full set of 113 galaxies
in the Frei et al. (1996) catalog (hereafter referred to as the Frei
sample). In contrast, C97 was limited to face-on spirals, or
ellipticals from the same sample; inclined systems, irregulars and
active galaxies were left out. 

\subsection{Photometry}

The current sample under study consists of 82 galaxies imaged in the
$B_J$ and $R$ bands (Gullixon \etal 1995), and 31 in the Gunn $g$,$r$,
and $i$ bands.  The scale for the $B_J$ and $R$ images is 1.34'' per
pixel with a field of view of 7' x 11'; for the $g$,$r$, and $i$
images the scale was 1.19'' pixel$^{-1}$ with a field of view of 16' x
16'.  The $g$,$r$, and $i$ images have relatively lower S/N than the
$B_J$ and $R$ images because of substantially shorter exposures
times. In addition to the usual processing (bias subtraction and
flat-fielding), Frei \etal removed foreground stars.  Occasionally, a
star that is projected on a bright part of the nucleus was not
removed, and causes a false high value for the asymmetry number for
those galaxies.

For consistency between the above photometric subsets, we use the \bv
\ colors listed in the {\it Third Reference Catalogue of Bright
Galaxies} (de Vaucouleurs \etal 1991, RC3) for purely photometric
purposes. (Structural parameters are computed from the images
described above.) The adopted \bv \ colors are {\it observed}, i.e.
uncorrected for Galactic and internal extinction, and heliocentric
velocity (the $k$-correction). These corrections are small and hence
have no qualitative impact on the conclusions of this paper.  We adopt
the distances and $B$ absolute magnitudes given by Tully (1988). We
note, however, that these M$_B$ include the above corrections. (A more
consistant treatment is presented in Jangren \etal 1999.) The adopted
values of M$_B$, $B-V$, and distance are listed in Tables 1 and 2.

\subsection{Sample Characteristics}

The Frei sample includes Hubble types from early ellipticals and S0s,
to late-type spirals, irregulars and galaxies with peculiar features.
Several of the galaxies in the sample contain features such as rings
and bars. Most of the galaxies in the sample are nearby, with a large
portion of the ellipticals coming from the Virgo cluster.  All of the
galaxies were chosen away from the galactic equator and are generally
near the Northern Galactic Cap.  However, the Frei sample consists
only of bright, high surface brightness galaxies. Hence, we are not
considering the large population of LSB or dwarf galaxies that make up
the bulk of all galaxies in the universe. It is important to realize
that the Frei sample is not likely to be an accurate representation of
the entire local galaxy population, but only samples a range of Hubble
types for the brightest, nearby galaxies.

For illustrative purposes, in the remainder of this paper we have
selected 21 galaxies that are representative examples of total sample
of 113 used in this study. These 21 galaxies span of T-types,
inclinations and colors. The remaining sample can be viewed at {\bf
http://astro.princeton.edu/$\sim$frei/galaxy}.

\section{The Asymmetry Algorithm}

The asymmetry A$_{\phi}$, where $\phi$ is the angle of rotation, is a
quantitative parameter based on comparing a source image with its
rotated counterpart. In principle A$_{\phi}$ is straight-forward and
quick to measure in its simplest form. We compare several different
methods, including those presented by Abraham \etal (1996a; hereafter A96)
 and C97 in the context of a $\phi = 180$\deg rotation, to determine if they 
are robust
morphological measures over a range of astronomical conditions.  On
this basis, we develop a more sophisticated approach, ultimately based
on the algorithm of A96.

\subsection{Image Preparation}
  
Before a rotational asymmetry measurement can be made, basic digital
image processing must be completed (i.e. bias subtraction, field
flattening, fringe correction, cosmic-ray removal, and bad pixel
interpolation). In addition, any objects in the galaxy's image which
are not part of the galaxy, such as foreground stars, or foreground
and background galaxies must be removed by blanking the contaminated
portion of the images, or subtracting the contaminating source. 
Contaminating sources tend to be at
random locations with respect to the primary source, and hence
residuals from this decontamination process will add to the asymmetry
depending on their amplitude relative to the source brightness. In the
case of unresolved contaminating sources, e.g. stars, the removal is
straightforward.  Foreground and background galaxies are more
difficult to remove or exclude both because of their more complex and
extended image structure, and because their identification as separate
from the primary source can require further information and
judgment. This complication becomes particularly important in
ultra-deep images such as the HDF, but we do not pursue this further
here as it is not an issue for the current study.

In addition to any foreground stellar images, the mean background
level for each image must be carefully subtracted. This will become
clear in the next section where we define the specific computational
algorithm.  A significant non-zero background level can radically
alter the measured asymmetry value. After a successful background
subtraction and object decontamination, however, it is then possible
to perform the asymmetry measurement without further image calibration
or processing.

\subsection{Methods of Computation}

The rotational asymmetry measurement procedure consists of (1)
defining an image center and ``extraction'' region; (2) rotating the
image about this center by angle $\phi$, and (3) performing a
comparison of the resultant rotated image with the original image,
within the extracted region. This comparison, however, can have
several mathematical forms. In published studies (e.g. C97, A96), the
comparison is made by subtracting the rotated image from the original,
although in principle the product or ratio might be useful. We have
found that only subtraction appears to work well, and discuss only
this possible algorithm further. The details of the centering, and
choice of rotation angle (\S3.3 and \S4.1 respectively) are critical,
but do not depend on the mathematical form of the asymmetry algorithm.

Once the rotated image is subtracted from the original image, we then
have a residual image of all asymmetric features of the galaxy.  The
asymmetry measurement is completed by measuring the amplitude of the
pixel values in the residual image, and normalizing this amplitude by
a corresponding measure in the original image.  Here again, several
mathematical possibilities exist.

The method used in C97 consists of summing the square of the pixels in
the residual image, and normalizing this value by dividing by the sum
of the square of the pixels in the original image.  Since the pixel
values in the residual image will be, on average, half negative and
half positive pixel values, half the above ratio gives gives the
asymmetry number for that galaxy: $$A_{rms}^2=\frac{\Sigma (I_o -
I_{\phi})^2}{2\Sigma I_{o}^2}$$ where $I_o$ is the intensity
distribution in the original image, and $I_{\phi}$ is the intensity
distribution in the image rotated by angle $\phi$.  The sum is over
all pixels within a prescribed, matching region of the original and
rotated images. We will call this procedure the ``RMS'' asymmetry
method.

Operationally similar is to replace the square by the absolute value,
as done by Schade \etal (1995) and A96.  Instead of
using a sum of squares of the original image, a sum of the absolute
value is used to normalize the residuals:
$$A_{abs}=\frac{\Sigma |(I_o - I_{\phi})|}{2\Sigma |I_{o}|}.$$ Again,
the sum is over all pixels within a prescribed, matching region of the
original and rotated images.  We will call this procedure the ``ABS''
asymmetry method.

For both methods the lowest possible value for the asymmetry parameter
is 0, while the highest is 1. A value of 0 corresponds to a galaxy
that is completely symmetric, that is the difference $(I_o -
I_{\phi}) = 0$ at all points in the difference image formed from
subtracting the rotated image from the original image. A values of 1
corresponds to a galaxy that is completely asymmetric such that on
average $|(I_o - I_{\phi})| = I_o$. Values for most galaxies lie
between these two extremes (see Tables 1, 2, 3, and 4 for
$\phi=90$\deg and 180\deg).

Comparing the above two asymmetry computations for $\phi = 180$\deg
rotations, we find that the values for A$_{abs}$ are similar to the
the values found for A$_{rms}$, as measured in the red. The A$_{abs}$
values have slightly larger asymmetries at high A as can be seen in
Figure 1 from the positive value of A$_{abs}$-A$_{rms}$ at large
A$_{rms}$. As a function of wavelength, the RMS asymmetries show a
clear trend towards higher differences in A(blue) - A(red) at larger
(red) asymmetries, as seen in Figure 2; this can be seen to a  lesser
extent for the ABS method in Figure 3. (Here, `blue' corresponds
either to the $B$ or $g$ bands, and `red' corresponds to either $R$ or
$r$ bands.) The average difference in A(blue) - A(red) for the RMS
method is 0.015 overall, and 0.05 at A(red)$>$ 0.1.  The ABS method
yields asymmetries that are fairly constant between bands with an
average A(blue) - A(red) of 0.005 over all asymmetry values.

However, we find the ABS method gives a tighter correlation between
asymmetry and color than the RMS method, contrary to our expectations.
The RMS method was expected to be a better indicator of star formation
since it weights higher the brighter, asymmetric features. The
brightness of a star forming region in a galaxy is a function of the
density, squared, and hence the RMS method should better trace higher
contribution from denser regions of star formation.  While either
method can be used to get physical information, the better correlation
of ABS asymmetries with color lead us to use the ABS method for the
remainder of the paper.

\subsection{Centering Corrections}
   
One of the most crucial aspects of the rotational asymmetry computation is
the choice of a center of rotation. Centers that differ by only a
small amount (relative to a galaxy's characteristic size) typically
produce substantially different asymmetry values. For example, a
change in the center of rotation by just one pixel for the Frei \etal
sample (roughly 1\% of a half-light radius) can change the value of
the asymmetry by as much as 50\%. However, this becomes less of a
problem when the scale of the galaxy becomes smaller, such that as the
sampling decreases so does the need for precise centering.

To overcome this centering problem, we define the center of
rotation to be the position which yields a minimum value for the
$\phi = 180$\deg asymmetry.  To find such a center in practice, an
initial guess is made for a galaxy's rotational center. This first step
can be automated by choosing the mean, or mode of the light
distribution as a reasonable initial guess; our tests indicate that
the initial guess does not alter the final asymmetry. In most cases,
there is a clear central pixel which is approximately in the center of
the galaxy, for example, the brightest pixel in the galaxy.  For
irregular and edge-on galaxies however, there is not a clear-cut
brightest point, or even a well defined center, and it is for these
galaxies that this method of minimum center is most effective.

Operationally, after the asymmetry is computed at that initial
position, the asymmetry is computed again for centers at the
surrounding eight points in a 3$\times$3 grid. The distance from the
central point to the eight surrounding points can be set at any
value. In this work, we use a distance of 0.1 pixels, corresponding to
roughly 0.1\% of the half-light radius. We use the task 'rotate' in
IRAF to perfom the asymmetry measurements via bi-linear interpolation.
If the asymmetry parameter at the center is lower than the asymmetry
value at any of the eight surrounding points, then the asymmetry
parameter is taken to be the value at the center.  The algorithm
effectively creates a synthetic grid of asymmetry values arranged by
x,y center positions.  If the center pixel does not give the asymmetry
minimum, as is usually the case, then the procedure repeats, with the
new center the pixel value where the minimum was located. This process
continues until a minimum is found for the asymmetry. Once this
location is found, we define this to be the `asymmetry centroid' and
use that computed asymmetry for the asymmetry parameter of the galaxy.

One possible problem with this method of finding the minimum asymmetry
is the existence of local asymmetry minima.  We have tested this by
computing the asymmetry parameter over a wide range of centers for a
set of 21 representative galaxies from the Frei sample. The second
columns of Figures 4, 5, 6, and 7 are pictorial representations of the
asymmetries values at all pixel locations about the inner 3x3 arcsec
of these 21 galaxies. It can be seen in these images that no
significant local minima exist throughout the image.  While the
detailed shape of these `asymmetry planes' differ from galaxy to
galaxy (and indeed, contain considerable information about the light
distribution) a well-defined minimum exists in each case. In the
presence of sufficient noise, this condition will break down. We
characterize this behavior via simulations in the following section.

\subsection{Noise Corrections}

The rotational asymmetry, as defined here, is by its very nature a
pixel-by-pixel difference algorithm, and hence can be substantially
affected by noise. Clearly this effect must be accounted for if
asymmetry is to be a robust classifier. An example of the effects of
noise on the asymmetry value of a galaxy is illustrated in Figure 8.
The effects of both correlated and uncorrelated noise are
relevant. Here, however, we develop a correction for uncorrelated
noise and defer handling of correlated noise to a later study on the
Hubble Deep Field.

The effects of uncorrelated noise in practice are easy to correct by
simultaneously performing the asymmetry measurement on both the source
and a neighboring, blank area of the image (see A96). The method for
computing the asymmetry of the `blank' area is the same as before,
with one exception: there is no normalization by the sum of the
original pixels, since in the case of the sky-subtracted background,
the sum is zero on average.  This procedure is then repeated in a
`centering' grid until a minimum of the noise is found, precisely the
same way the minimum asymmetry of the object is found. This `blank'
asymmetry value is then subtracted from the value measured for the
object, thereby correcting statistically for the effects of
uncorrelated noise present in the object image.  The final formula
used to compute the asymmetry can be written as: $$ A_{abs}= {\rm min}
[\frac{\Sigma |(I_o - I_{\phi})|}{\Sigma |I_{o}|}] - {\rm min}
[\frac{\Sigma |(B_o - B_{\phi})|}{\Sigma |I_{o}|}],$$ where $I$
represents the image pixels, and $B$ represents the blank region
(background) pixels.  Note that the possible range of asymmetry values
now spans from 0 to 2. However, due to the application of the
minimization condition, the values are rarely ever greater than 1, and
this primarily occurs when $\phi = 90$\deg.

When defining a blank region it is necessary to
either use an extraction region the same size as the one used on the
galaxy, or more practically, scale the sum of the blank region by the
relative size of the object to `blank region' areas.  The extraction
region used to define the background should be big enough to be
representative, but should be small and distant enough from the galaxy
so as to not include any diffuse light -- and hence gradients -- from
the sources.  Figure 8 shows the differences between the computed
asymmetry of a galaxy with added noise and the original asymmetry as a
function of S/N.  It can be seen from this plot that the asymmetry
differences becomes very large at lower S/N.

To test how successful our algorithm of removing noise is at
effectively reproducing the correct value for the asymmetry parameter,
we artificially degraded the Frei \etal galaxy images by adding
simulated, random noise. The noise-corrected asymmetry values measured
in these images allowed us to assess the systematic behavior with
S/N. We compute the S/N as ratio of the signal from the galaxy within
the half-light radius to the noise contribution from the sky, source,
and detector read-noise within the same aperture.  Results of these
S/N tests are shown in Figure 9. We use 35 galaxies from our sample
which span all Hubble types and inclinations to perform these
simulations. We find that for all these various galaxies, the value of
asymmetry parameter found at lower S/N is, on average, still near the
values found in the original images.

Below S/N $\sim$ 100, the noise begins to heavily dominate the
asymmetry. In this regime, the rotation center yielding the minimum
asymmetry (described in Section 3.3) is determined largely by the
noise field. This is compensated in the noise correction since we find
the minimum again for the blank field where the background correction
is calculated. If we did not re-center on the blank region when
finding the background-level asymmetry, this correction would get
relatively larger than the galaxy's asymmetry at successively lower
S/N values. As a consequence, the corrected asymmetry would
systematically become negative at low S/N values. In our current
scheme, we avoid this pit-fall. Even for S/N $<$ 100, our simulations
show that measured asymmetries have error bars that still overlap with
the actual value, although the errors are very large.  Errors for
images with S/N $>$ 500 are typically around 0.02 (rms), while at S/N
between 100 and 300 have errors around 0.05 (rms). At S/N $<$ 100, the
error on the asymmetry become very high, i.e. exceeding 0.60 (rms) for
S/N $<$ 50. From Figure 9 however, we conclude that for all galaxies
with S/N values $>$ 100 their asymmetries can be computed reliably,
which we define to be when the rms errors are less than 0.05.

A feature of our asymmetry algorithm, which comes naturally from the
noise corrections, is an ability to estimate an error on the computed,
corrected asymmetries. We have tested this via Monte Carlo methods and
find that these estimates are accurate.\footnote{The programs used in this 
paper to compute the various asymmetries can be obtained by emailing one of
the authors.}

\section{Results}

The following presentation is based on the results in Tables 1 and 2,
which list the asymmetry values and their estimated errors for the 113
galaxies in the Lowell and Palomar sample for $\phi$ = 180\deg and 90\deg,
computed within the radius r($\eta$=0.2), and calculated as described
in the previous section with centering and noise corrections. The
extraction aperture based on the $\eta$-function is described in
section \S4.2.
  
\subsection{Asymmetry as a function of $\phi$}

Rotational asymmetry, as we have defined it, has heretofore been
explored only in the context of 180\deg rotation. However, rotation by
other angles can potentially yield more physical and morphological
information.  In particular, the azimuthal variations of galaxy light
profiles can be probed, akin to the seminal work of Schweizer (1976a,
1976b). While it is not our intent to pursue such a detailed study
here, we do show that other angles of rotation can provide diagnostics
which allow us to improve upon the utility of the 180\deg asymmetry
correlation with color.

\subsubsection{180\deg Rotation: flocculent and dynamical asymmetries} 
      
The rotation of a galaxy by 180\deg should yield the minimum
rotational asymmetry since most galaxies appear to have a strong
azimuthal axi-symmetry. As such, A$_{180}$ is expected to be sensitive
both to either large scale departures from axi-symmetry (e.g. dynamical
disturbances or asymmetric modes), or small scale departures from
axi-symmetry in the form of star-forming regions. For normal galaxies,
it is the later which is of particular interest.

Both ABS and RMS asymmetry methods clearly show that the $B_J$ band
almost always has a higher asymmetry value than the $R$ band (Figures
2 and 3). This is consistent with the shorter wavelength $B_J$ band
sampling more sensitively the light from the younger stars which are
distributed non-uniformly throughout the galaxy. As seen in C97, the
difference between A($B$) and A($R$) increases at higher values of the
asymmetry. Jangren \etal (1999) also show that the difference between
A($B$) and A($R$) increases for galaxies with bluer colors. These
trends are a further indication that recent star formation is the cause
of asymmetries, and in particular that, for the most part,
contributions to the asymmetry of normal galaxies come from blue star
forming regions in the arms of spirals or irregular galaxies.  We
refer to this as `flocculent' asymmetry.  The vast majority of the
sample have `flocculent' asymmetries and follow the color-asymmetry
sequence as seen in Figure 10.

However, asymmetries can also be caused by dynamical events, such as
the interaction or merging of two galaxies.  We call the asymmetry
from these interactions `dynamical asymmetry.' A dynamical event can
warp and extend a galaxy, deviating its structure from the symmetric
`ground state.'  Dynamical asymmetries add to the flocculent asymmetry
such that dynamically distorted galaxies have asymmetries which are
always higher than asymmetries caused solely from star formation
processes. The galaxies in the Frei sample that are in the process of
a galaxy interaction/merger are labeled in Figure 10; they are too
asymmetric for their colors. We will explore this issue, its relation
to other physical features of galaxies, and how to effectively use it
for morphological classification in later sections.

\subsubsection{90\deg Rotation: Ellipticity}

The asymmetry in a 90\deg rotation (A$_{90}$) is almost always larger
than a 180\deg rotational asymmetry, as seen in Figure 11.  The median
ratio A$_{90}$/A$_{180}$ for the Frei \etal sample is roughly 4, but
has a wide range at low A$_{180}$.  There also is a substantial range
in A$_{90}$ for galaxies spanning all values of 180\deg asymmetry. We
interpret these observations simply to mean that galaxies have a
strong 180\deg symmetry yet a variety of projected shapes, which leads
to a wide range of 90\deg rotational asymmetry.

Indeed, A$_{90}$ can be used to estimate the ellipticity of the light
profile. For most normal galaxies A$_{90}$ correlates strongly with
galaxy axis ratio, b/a, as seen in Figure 13.  This can also be
verified by visual inspection of Figures 4-7, which are sorted (for
each type) by A$_{90}$. There is, however, a small contribution of
from flocculence to A$_{90}$. This can be seen by plotting the axis
ratio instead against the difference of A$_{90}$ and A$_{180}$ (Figure
13, bottom panel). A tighter correlation can be seen between A$_{90}$
and the axis ratio in the bottom panel after A$_{90}$ has been
`corrected' for the flocculent component measured by A$_{180}$. To
first order A$_{180}$ and A$_{90}$ should have comparable amplitude
due to flocculence alone. While we tried various mathematical schemes
to correct A$_{90}$ using A$_{180}$, simple subtraction worked best.
On this basis, we suggest that the quantity A$_{90}$ - A$_{180}$ can
be used in place of directly measured axial ratios, for statistical
purposes. This is particularly useful for distorted galaxies without
well-ordered, elliptical isophotes; in this case the isophotal
measurement of ellipticity is problematic, and A$_{90}$ may prove a
useful substitute or diagnostic via comparison to standard isophotal
techniques.

As a result of A$_{90}$'s sensitivity to the global azimuthal shape of
the light profile, A$_{90}$ does not correlate as well with
morphological type or color as does A$_{180}$. For example, at each
T-type, there is a larger scatter in the values for A$_{90}$ than for
values of A$_{180}$. A$_{90}$ measured in two different wavebands,
e.g. $R$ and $B$ for example, do correlate strongly (Figure 12, both
panels). But while A$_{90}$ in the $B_J$ band is a bit more
asymmetric, we do not see as strong an increase in the difference
between A(B)$_{90}$ - A(R)$_{90}$ as we do for A(B)$_{180}$ -
A(R)$_{180}$ (cf Figure 3 and the bottom panel of Figure 12). We can
infer from these observations that A$_{90}$ is not particularly
sensitive to the flocculent asymmetry due to star-formation that
A$_{180}$ most effectively measures.  We do note that the dispersions
in A(blue)-A(red) are about twice as large for A$_{90}$ than for
A$_{180}$. This may be attributable simply to the larger A$_{90}$
values.

It is worth noting that spheroidal systems (ellipticals and S0s) stand
out as never having A$_{90}>0.70$ -- uncharacteristic of the other
T-types. We expect all ellipticals, being largely devoid of
star-formation, to have A$_{180}$ values $\approx$ 0 along the entire
elliptical sequence from E0 to E7. However, only round elliptical
galaxies (E0) would be expected to have zero asymmetry values for any
rotation angle $\phi$. The more the elliptical deviates from a round
shape, the more the value of A$_{\phi}$ will change as the angle of
rotation $\phi$ changes. Since a measure of A$_{90}$ can give an idea
of the shape, hence the morphology for elliptical galaxies can also be
estimated.  There is a slight correlation between the Hubble
morphological type index for ellipticals (e.g. E6, where 6 = 10 * (1 -
b/a)), but we do not have enough ellipticals in our sample to say
whether A$_{90}$ is useful for determine the Hubble sub-type for
ellipticals. As is always true for determining elliptical sub-type,
projection effects are an issue.

It also is interesting to note that there is some scatter in the
A$_{90}$ values even for E0s. Either E0s are not completely round, or
they contain a hidden disk structure, as proposed by Kormendy \&
Bender (1996) to explain the isophotal features of elliptical
galaxies.

A final point to note about Figure 11 is the objects with extreme
90\deg asymmetry values, i.e. A$_{90} > 0.8$ are almost exclusively
edge-on systems. Hence by using both A$_{180}$ and A$_{90}$, these
galaxies with high inclinations can be singled out in an automatic
process.

\subsubsection{Azimuthal Rotation Profiles}
 
To test if other angles are instructive as morphological indicators,
we have computed the asymmetry index A$_{\phi}$ for rotation angles
$\phi$ from 0 to 360\deg with 1\deg increments for the 21 galaxy
subset representative of the Frei sample defined in \S2. These are
displayed in the right-most columns of Figures 4-7. Naim \& Lahav
(1997) have also considered a variety of rotation angles in the
context of attempting to define galaxy `peculiarity.' The treatment
here is somewhat more general.

The salient feature of all the asymmetry-rotation angle profiles is
the remarkable similarity of the basic shape: the asymmetry profile
(which starts at 0 by definition at $\phi=0$\deg) rises steeply at
first, plateaus near 90\deg, where there is a maximum, and then
descends to a local minimum at 180\deg.  To first order, these
profiles are remarkably similar for counter-clockwise rotation from
$0<\phi<180$ and clockwise rotation from $0>\phi>-180$. In Figures
4-7, the counter-clockwise and clockwise rotation profiles through
$0<|\phi|<180$ are shown as solid and dashed curves respectively. The
fact that dashed curves are frequently {\it not} seen reflects this
striking similarity.  To second order, these profiles are also remarkably
similar even about 90\deg symmetry points, i.e. the curves from
$0<\phi<90$ and $90<\phi<180$ are often indistinguishable except near
$\phi=0$ and $\phi=180$. The profiles are folded every 90\deg in
Figures 4-7 to illustrate this point. We find, like Kornreich \etal
(1998) that most galaxies are symmetric in $\phi$.

The location of the local minimum of the asymmetry rotation profile at
the 180\deg has some interesting implications.  For example, if a
galaxy has a triple arm pattern, or a four arm pattern, this could be
reflected in minima at 120\deg or 90\deg, respectively.  Previous work
on this topic found a significant number of galaxies with triple arm
patterns, revealed by rotations and subtractions similar to what is
presented here (Elmegreen \etal 1992). While one might conclude that
such symmetries must be rare, it is more likely that they are simply
of lower amplitude than the basic axi-symmetry inherent to virtually
all galaxies in this sample. This axi-symmetry, coupled with
projection, determine the primary features in the rotation profiles as
described above.

As previously discussed, one may infer that galaxies are axi-symmetric
at $180$ rotation angles because they are dynamically relaxed;
axi-symmetry is the `lowest energy state.' Perturbations from this
`ground state' due to recent dynamical events, however, should be
revealed in the details of the asymmetry rotation profiles.  Indeed,
in detail, there {\it are} variations in the slope of these profiles
at all $\phi$. These variations appear to correlate with galaxy type.

For example, there are variations in both the location of the maxima
near $\phi=\pm90$ and local minima near $\phi=\pm180$. The maxima tend
towards smaller $\phi$ for latter types, qualitatively consistent with
the results of Naim and Lahav (1997) for what they describe as
increasingly `peculiar' systems. Likewise, there are a variety of
slopes at any given value of $\phi$, and many cases where there are
subtle variations in the basic 90\deg symmetry of the rotation
profiles. Measurements of the steepness of slope at several fiducials
(e.g. 0\deg, 30\deg, 60\deg, and 90\deg) could reveal interesting
correlations with other physical parameters.  While these differences
could be quite illuminating for the physical understanding and
classification of these systems, such an analysis is left for future
work.

\subsection{Symmetry as a function of Galactic Radius}

The asymmetry index is highly sensitive to the aperture size, as
illustrated in the third panel of Figures 4-7.  The question is, then:
what radius should be used? While in principle any radius can be used,
one well-defined alternative is to choose a single radius tied to the
physical scale of the galaxy. Clearly whatever choice is made, it must
be used {\it consistently} for comparative purposes. We approached
this decision by computing A$_{180}$ first for a wide variety of
different, well-defined radii tied to the metric (physical) scale of
each galaxy (as described below).  We then determined empirically
which asymmetry value correlated best with other physical parameters,
such as color.

Traditionally, galaxy radii have been defined using a surface
brightness criteria, such as the Holmberg radius at $\mu_{B} = 26.5$
mag arcsec$^{-2}$. These radii, defined to be at the galaxy isophote
where the galaxy is just distinguished from the sky on a particular
set of photographic plates, are often inadequate, inappropriate, or
ill-defined when trying to compare a wider range of galaxy types at
various distances.  For example, low surface-brightness galaxies pose
a particular problem here. The $(1+z)^{-3}$ surface intensity dimming
(detected photons per band-pass), coupled with $k$-corrections,
further makes the definition of an isophotal radius difficult at best,
even for normal galaxies. Evolution only adds to the complication in
the isophotal approach; and photometric zero-points are also of
concern. In short, isophotal radii, while operationally convenient to
measure, are prone to a wide variety of systematics and therefore are
among the least robust measures of galaxy radii.

When establishing a characteristic galaxy size, a radius should be
used that is independent of the over-all normalization of the
surface-brightness distribution, and hence independent of redshift and
photometric calibration. Two possible alternative to defining radii,
$\theta$, are based on (i) the $\eta$-function (Petrosian 1976), and
(ii) the curve of growth, g($\theta$). The latter requires a robust,
working definition of a total magnitude. Such schemes have been
discussed at length elsewhere (e.g. Bershady \etal 1994, Bershady,
Lowenthal \& Koo 1998), and here we adopt the protocol developed by
Jangren \etal (1999), which uses twice the radius where $\eta = 0.2$
to define the total light.\footnote{We follow Kron's (1995) suggest to
use the inverted form of $\eta (r) \equiv I(r)/\langle I(r)\rangle$,
which equals one at the center of the galaxy and approaches zero at
large galactic radii, $r$.}  Tables 3 and 4 list the asymmetries
computed at the various $\eta$ and curve of growth radii. Based on
these measures, we have determined that the radius where $\eta=0.2$
delivers the best correlation with $(B-V)$ color.  Asymmetries
computed with the other radii also show a correlation in color.  In
principle, any radius can be used, as long as it is consistently used
when comparing different galaxies. Occasionally, the $\eta=0.2$ radius
yields extraction regions which extend beyond the image boundaries for
the Frei sample. When this occurs, we compute the asymmetry at the
largest radius possible.

The trend of asymmetry versus aperture radius is useful to determine
where sources of asymmetry in a galaxy are coming from, e.g.
different relative amounts of nuclear and disk star-formation, or
low-frequency assymetric structures such as off-center bars and oval
disks.  It is immediately obvious from inspection of Figures 4-7 that
for no galaxy is the asymmetry a constant function of radius. This
shows the importance of using a standard radius to compute the
asymmetry for a galaxy.
  
We are able to make a few general observations about the distribution
of asymmetric light in a galaxy, which is clearly a function of
T-type. Ellipticals and S0s look very similar, with their asymmetry
peaking at very low radii, and modestly decreasing outward, and then
increasing rapidly at high very large radii (Figure 4). The central
peak may be due to structures commonly found in the centers of
ellipticals (Lauer \etal 1995.) The sharp rise in asymmetry at large
radii is not necessarily a physical effect, but is a result of the
noise correction in a regime where the noise dominates the signal.


In contrast, spiral galaxies (Figure 5 and 6) show an increase in
asymmetry with radius, sometimes with local or global maxima at
intermediate radii.  For the early disks (Sa-b, Figure 5), the maximum
is usually at or near the half-light radii. The late-type disks (Sc-d,
Figure 6) have their peak asymmetry values at higher radii, usually
beyond the half-light radii.  The irregulars (Figure 7) also show a
remarkable increase in asymmetry as a function of radius, with the
inner parts comparatively symmetric.
Maxima in the asymmetry as a function of radius diagrams (Figures 5
and 6) for disk systems are typically at the locations of HII regions
or dust lanes in spiral arms.

To summarize the observed trends: as galaxies become systematically
later in type, the amplitude of asymmetry increases {\it and} the
increase with radius becomes stronger. That is, the latest type
galaxies are pronouncedly more asymmetric in their outer parts than in
their inner parts compared to earlier-type galaxies.  This
characteristic could in principle be an objective method for
determining a galaxy's Hubble type.  One physical explanation for this
trend of larger radii peaks in asymmetries for later types is the
increasing dominance of actively star-forming disks in late-type
systems. However, other possibilities include increasing specific
star-formation rates with galacto-centric radius (Hodge and Kennicutt
1983; Ferguson \etal 1998); the late-time accretion of material in the
outer disk with a dynamical-relaxed core (e.g. as proposed by Rix \&
Zaritsky 1995); or tidal interactions with nearby galaxies. These
effects, however, would have to be correlated with morphological type.

A more sensitive probe of the physical cause of these radial trends in
asymmetry would be to examine the rotational asymmetry in annuli and
to compare the `specific asymmetry,' as measured in these annuli, to
the `specific color' of the annuli. As we have argued above, a
significant component of the color-asymmetry correlation is due to
what we have termed flocculent asymmetry, i.e. the irregular
distribution of HII regions in an otherwise axi-symmetric
system. However we have also noted that departures from the
color-asymmetry trend for normal galaxies appears to be an indication
of larger scale asymmetry -- what we refer to as dynamical
asymmetry. These statements describe the relation between the
integrated asymmetry and color of galaxies. In analogy, departures
from the color-asymmetry trend in a given annulus might help
disentangle where and when a galaxy's asymmetry is dominated by
star-formation or large-scale dynamical asymmetries. The analysis of
Kornreich \etal (1998), by excluding the inner portion of the galaxy,
achieves this goal. Using rotation instead of a modal analysis, and
adding more radial resolution would offer further dimension to this
study, and we will pursue this elsewhere.

\subsection{The Correlation of Color and Asymmetry}

The color-symmetry diagram is a powerful tool that can be used for
both morphological and physical differentiations of galaxies, and
hence can be used as a diagnostic for understanding distant
galaxies, and galaxy evolution (C97; Conselice \& Bershady 1999). The
color-asymmetry diagram (Figure 10), when using only the normal face
on galaxies presents a tight, linear sequence, as shown in C97. This
normal-galaxy sequence represents a lower limit where, for any given
asymmetry, there are no galaxies with bluer colors; we refer
to this as the 'fiducial' galaxy color-asymmetry sequence.

When the entire Frei sample is included (with normal, active, edge-on,
irregular and peculiar galaxies), still no galaxies are seen to occupy
the symmetric-blue region (bottom left part of Figure 10). This shows
that blue galaxies -- at least in the Frei \etal sample -- are
asymmetric galaxies. Symmetric blue galaxies may exist, however, at
higher redshift (e.g. Schade \etal 1995).
  
To first order, the color-asymmetry diagram of Figure 10 gives a
wealth of diagnostic information for classifying galaxies of different
morphologies. The elliptical and spirals separate distinctly in this
diagram. Ellipticals, denoted as large circles, are always found at
the upper left portion of the digram where red and symmetric objects
are located.  Edge on systems, located by tiny dots, occupy the space
to the right of the diagram. The spirals and irregular galaxies are
the objects plotted as boxes (early types Sa-b), as stars (later
types, Sc-d), and as open circles (irregulars) are always bluer and
more asymmetric than the ellipticals. In short, the late type galaxies
versus early types can can clearly be differentiated in an automatic
method via the color-asymmetry diagram.

In addition, the color-asymmetry diagram presents the ability to
single out galaxies undergoing an interaction or merger. In Figure 10,
we have labeled the galaxies which are most likely undergoing an
interaction with another galaxy.  These objects stand out as being
{\em too asymmetric for their colors} and therefore, star formation
can not be the cause of the asymmetry, as it is for the galaxies on
the fiducial sequence.  Of the six galaxies listed, only one (NGC
5792) is an early type, the remainder are Sc-d systems.  NGC 5792's
high asymmetry can be accounted for in part by the residuals left
after subtracting a very bright star near the center.  One galaxy, NGC
4088 (Arp 18) has an outer arm receding and an elongated nucleus
(Vorontsov-Velyaminov 1977; Dahari 1985) -- possible evidence for a
recent tidal interaction. NGC 4254 is a Virgo cluster sprial that is thought 
to have an
external driving mechanism (Rauscher 1995) -- possibly a result of
tidal interactions with the core of the Virgo cluster. These are the
only galaxies in the Frei sample that have evidence for interactions,
and they all can be distinguished by examining their place in the
color-asymmetry diagram. Other galaxies that have similar colors as
interacting galaxies, such as the Magellanic Irregular NGC 4449, fall
along the fiducial sequence (Figure 10) as its asymmetry is
caused by star formation.  When applied to images of distant galaxies,
this segregation can be used as an effective tool for disentangling
possible evolutionary effects, e.g. the role of mergers and
interactions as opposed to star formation.

The edge-on galaxy systems stand out in Figure 10 as objects that are
too asymmetric for their colors. However, they generally are not as
asymmetric as the interacting galaxies, and have redder colors.
Indeed, a large portion of the scatter of the galaxies in the
color-asymmetry diagram can be accounted for by this inclination, as
demonstrated in Figure 14. Here the color-asymmetry diagram is
revisited but with galaxies of higher inclinations labeled with larger
symbols. (The four sizes represent quartile bins in the observed
$R$-band A$_{90}$ distribution.) The galaxies that contribute to the
scatter the most also are the most inclined, almost without
exception. As one might infer from the tight correlation in Figure 13,
one reaches the same conclusion marking sources according to axis
ratio instead of A$_{90}$. This raises the possibility that the
scatter could be reduced with a hybrid asymmetry-color diagram which
included or corrected for inclination.

Since inclination potentially can effect both the apparent asymmetry
and color, a first step is to determine how each of these parameters
is being affected.  That galaxies become redder with increasing
inclination is of little doubt, yet the behavior of asymmetry with
inclination is less obvious. When we plot the color-asymmetry
diagram using colors corrected for extinction (de Vaucouleurs 1991)
the scatter is dramatically reduced, as seen in Figure 15. From this
exercise one may conclude that the effects of inclination on
asymmetry are, in comparison to color, second order.

While inclination-corrected colors yield a stronger asymmetry-color
relation, ideally we would like a method of correcting for inclination
that is robust over a range of distance.  Our concern is that the
inclination-corrections of de Vaucouleurs (1991) represents an
excellent, but fine-tuned algorithm based on large, normal galaxies,
and require an accurate identification of a galaxy's T type and b/a at
a rest-frame isophote comparable to what is used locally. We already
have discussed at length the problems of basing measurements on
isophotes. Our hope is that A$_{90}$ may offer a more robust
substitute. However, examining correlations of the residuals about the
observed color-asymmetry relation with A$_{90}$, for example, there is
no simple empirical way to correct these residuals. This is obvious
from a more careful inspection of Figure 14: while it is true that
most of the outliers from the fiducial sequence have large
A$_{90}$, a good number of galaxies with large A$_{90}$ also lie in
this fiducial sequence.  We suspect that more information about
type -- possibly using light concentration -- -- may allow us to
achieve a better inclination correction. To develop this possibility
requires a larger training sample and will be explored in a future
paper.

\subsection{The Correlation of Asymmetry with Other Morphological and 
Physical Parameters}

\subsubsection{Asymmetry and Image Concentration}

There is also a correlation between asymmetry (A) and the
concentration (C) of light for galaxies, as first found by A96, and 
revisited here in Figure 16 using new measurement
algorithms. The concentration index is from Jangren \etal (1999), and
is defined as the logarithm of the ratio of the radii enclosing 20\%
and 80\% of the light.  In principle, this diagram could be used for
the same classification purposes as color and asymmetry. It could be
argued that asymmetry and image concentration are methodologically
superior since, unlike color, measures of both asymmetry and
concentration do not require knowledge of the source redshift -- at
least to first order.  However, as discussed by Jangren \etal (1999),
there are substantial dependencies of these image structural
parameters on redshift which must be corrected, thereby mitigating their
putative advantage. Our A-C diagram, which has a tighter correlation
than A96, is as tight as A-color. One difference is
that the A-C is double valued. Nonetheless, this can be handled with
an affine parameter, as A96 has effectively done.
An advantage that A-C has over A-color is that the edge-on galaxies do
not appear to have a substantially different distribution. This is
another indication that A is only weakly affected by inclination.

\subsubsection{Asymmetry, T-type, and Luminosity}

As was shown in C97, the asymmetry value for a galaxy correlates with
T-type. For the larger sample used in this paper, the asymmetry -
T-type plot is shown in Figure 17, with a similar result, namely that
the Hubble sequence is one of increasing asymmetry, with a substantial 
scatter.

C97 also showed that asymmetry does not correlate strongly with
absolute magnitude, illustrated here in the bottom panel of Figure 18
for the larger sample. One might expect such a correlation since
color-luminosity relations are known to exist for all galaxies even in
the optical (Huchra 1977, Bershady 1995), and color and asymmetry are
tightly correlated.  However, we note that given the limited dynamic
range in luminosity of this sample, the lack of a strong correlation
between asymmetry and luminosity is not surprising. For example, the
color-luminosity correlation exhibited by this sample is paltry for
this reason, as illustrated in Figure 18a.

When we test the relationship between the asymmetry parameters and the
van den Bergh luminosity classification (van den Bergh 1960a, 1960b)
for the subsample with known van den Bergh luminosity types, we again
do not see a correlation between these two morphological
parameters. This suggests that asymmetry is a perpendicular
morphological parameter to the luminosity class of a galaxy. A
detailed system of morphology using the T-type, asymmetry and
luminosity class of a galaxy is possible, and will be the subject of a
separate paper.

\section{Application to Distant Galaxies}

In \S3.4 we established that asymmetry can be well-measured for S/N
$\geq$ 100 (within the half-light radius). Here we determine the other
most relevant observational parameter: the required spatial resolution
for accurate asymmetry measurements. For this purpose, we define
$\epsilon$ as the ratio of the angular diameter subtending 0.5
h$_{75}^{-1}$ kpc at a given distance ($\theta_{0.5kpc}$, where
h$_{75}^{-1}$ = H$_0$ / 75 km s$^{-1}$ Mpc$^{-1}$) to the angular
resolution of the image ($\theta_{res}$): $$\epsilon \equiv
\frac{\theta_{0.5kpc}}{\theta_{res}}.$$ The choice of numerator stems
for the results of our simulations, which, as we will discuss, show
the asymmetry parameter was found to be recoverable when the
resolution element was greater than 0.5 h$_{75}^{-1}$ kpc.  Hence
$\epsilon$ provides an index of the resolving power of a telescope
relevant to the measurement of a galaxy's asymmetry.

For subjective morphological classifications, such as the classical
Hubble scheme, as one might expect that as long as a galaxy's large
scale structure, (e.g. the bulge and disk components) can be resolved,
at least a rudimentary classification can be given. For large
galaxies, one might estimate that $\sim$ 1 kpc physical resolution is
marginally adequate, i.e. $\epsilon \sim 0.5$.  This resolution is
afforded by HST essentially at all redshifts.  However, since the
appearance of the spiral arms and the flocculence of the disk is
important in the Hubble classification, higher spatial resolution
($\epsilon \geq 1$) is required to make classifications comparable to
what is done for nearby galaxies even from ground-based
images. Nonetheless, the Hubble scheme has been used to classify
distant galaxies as seen in moderate to deep Hubble Space Telescope
images (van den Bergh \etal 1996), and has been automated using neural
networks (e.g. Odewahn \etal 1996).

Since a critical component of asymmetry in normal galaxies also is due
to `flocculence,' then there could be substantial systematics of the
observed asymmetry with resolution (distance).  In C97, a slight
distance effect was seen: the value of the asymmetry parameter (A) was
found to decrease very slightly with increasing distance, i.e.  the
sources appeared to become more symmetric. We find a similar type of
relationship using the entire Frei sample, where galaxies on average
become more symmetric at higher distances. Mitigating this effect is
that the most distant sources in our sample tend to be red
and hence have lower intrinsic asymmetry.

To determine the importance of image resolution on the measured
asymmetry, we simulated the appearance of the Frei \etal galaxy
sample, as they would be observed at large distances with, e.g., HST.
The simulations are simple in that we considered only the change in
apparent size relative to a fixed apparent pixel size. The effects of
image blur and redshift (i.e. change in observed portion of the galaxy
rest-frame spectrum) were ignored. In other words, we assume here that
the point-spread-function (PSF) abberations are small compared to the
pixel size, and that the observed band-pass is shifted with redshift,
respectively. Both of these assumptions are reasonable for
substantial data sets combining multi-band images using the HST
Wide-Field Camera-2.

The sources in the simulated `degraded' images had their asymmetry
computed by the same method described in Section 3.  The centering
algorithm is particularly important here. For less well-resolved
sources the relative error in selection of an image can be large. We
observed that the mean and mode of the light distribution changed
significantly at the pixel level as galaxy's image was sampled more
coarsely, even though the original images of most galaxies had a
visually well-defined center.

However, even using our routines which find the minimum asymmetry, we
find that almost all galaxies decrease in their asymmetry value when
artificially degraded, as shown in Figure 19.  Interestingly, we find
that the most symmetric galaxies (ellipticals) have asymmetries that
initially {\it increase} at coarser sampling redshifts, an effect also
noticed by Wu, Faber \& Lauer (1997). Recall that the higher the value
of $\epsilon$, the better a galaxy is spatially resolved and
sampled. As $\epsilon$ decreases, so too does the asymmetry
parameter. However, while the asymmetry parameter does decrease in
Figure 19, the $\epsilon$ values change by a factor of twenty; for
$\epsilon >1$, the measured asymmetries remain close to the
high-$\epsilon$ value.

Fortunately, the angular size of a galaxy changes little beyond $z
\sim 0.7$ for a wide range of cosmologies.  At $z = 1.25$, a galaxy
observed in the HDF, where $\theta_{res} \sim 0.045$, will have
$\epsilon \sim 1.3$ (q$_0=0.5$, $\Lambda=0$), and greater at higher
and lower redshifts for this cosmology. In comparison, a typical Frei
\etal sample galaxy has a size of about 4', with a pixel size of
1.35'', with $\epsilon = 5$. These $\epsilon$ values are both in a
range where asymmetry changes little with $\epsilon$; hence their
asymmetries can be reliably compared. We nominally confirm, then, the
results of A96 that resolution degradation of
asymmetry for distant galaxies observed with HST is not a significant
effect {\it if the highest available resolutions are obtained}. We
note, however, that $\epsilon$ for distant galaxies in the HDF is just
on the edge of being acceptable for asymmetry measurements. For q$_0 <
0.5$ $\epsilon$ remains above 1, but for coarser pixel sampling
$\epsilon$ quickly falls below 1 -- an issue particularly germane to
NICMOS imaging data unless it is properly over-sampled.

Our computations of how the angular diameter will change as a function
of redshift, illustrated in Figure 20, shows that galaxies at any
redshift imaged with the {\em Hubble Space Telescope} can have their
asymmetries reliably computed with the suitable instrument. Even with
excellent seeing on ground-based telescopes (e.g. WIYN 3.5m telescope
in 0.6'' seeing), asymmetry measurements can only be computed reliably
out to z$\approx$0.1. This shows the importance of instruments like
{\em HST} and the {\em NGST} for morphological studies of high
redshift galaxies. Similarly, high-order adaptive optics on large
ground-based telescope will permit asymmetry measurements at large
distances.

A caveat worth noting is that our simulations are based only on large,
luminous galaxies. There exists the possibility that physically small
galaxies have a systematically different power spectrum of luminosity
fluctuations due to a scale-dependence on the number of large
star-forming sites (Elmegreen \& Efremov 1996). If so, our prescription
based on a single $\epsilon$ index may be overly simplistic.

\section{Conclusions}

We have describe the use of the 180\deg rotational asymmetry parameter
for both physical and morphological diagnostic uses, and placed its
computation on a firm basis to allow comparisons between nearby and
high redshift galaxies. We tested several methods of computing
rotational asymmetry, concluding that a substantially modified version
of Abraham \etal's (1996a) method correlated best with physical
parameters such as color.  The three critical modifications which we
have developed here include: (1) an unambiguous and robust definition
of a center of rotation; (2) a new method for correcting for noise
which uses the same minimization method that is independently applied
to the source; and (3) a well-defined radius within which asymmetry is
measured.  We artificially redshift galaxies to determine the
dependence of our asymmetry measure on resolution and S/N asymmetry
measurements.  As expected, distance effects a galaxy's asymmetry
measurement by making the galaxy more symmetric. With this revised
rotational asymmetry algorithm, we estimated via simulations that
asymmetry can be robustly measured down to an integrated S/N of 100
(as measured within the half-light radius), and with spatial
resolution above 0.5 h$_{75}^{-1}$ kpc.

We also investigated the asymmetries computed as a function of the
rotation angle, $\phi$, finding a strong minimum in asymmetries at a
rotation angle of 180\deg, and maxima near 90\deg. To first order,
these asymmetry rotation profiles are remarkably similar from galaxy
to galaxy. To higher order, variations in these profiles undoubtedly
offer further information for probing the light distributions of
galaxies.  We also find different behavior in the asymmetry of a
galaxy as a function of its radius. Elliptical galaxies have a modest
decline in asymmetry with increasing radius, while later type galaxies
have a pronounced, and opposite trend.

We also find that 180\deg asymmetries correlate well with Hubble
morphological type, color, and concentration.  We suggest that the
color-asymmetry relation for galaxies is a fundamental one that can be
exploited in several different ways to obtain information about
galaxies.  From the color-asymmetry diagram alone, we are able to
distinguish between spirals, ellipticals, edge-on galaxies, as well as
interacting galaxies. The 90\deg asymmetries correlate strongly with
ellipticity (b/a), and can be used to tighten further the correlation
of color to asymmetry.

We find no correlation between the absolute magnitude or van den Bergh
(1960) luminosity class and the asymmetry parameter of a galaxy.
Asymmetry, which appears to be closely related to color, and hence the
relative youth of a galaxy's stellar population, appears to be a
morphological indicator perpendicular to this luminosity class. While
the degree of asymmetry represents an indicator roughly parallel to
the Hubble sequence, our expectation is that asymmetry can be used in
conjunction with other quantitative parameters to develop a new,
refined morphological classification. Such a classification would have
a more directly interpretable physical basis, but need not -- and
indeed, should not -- be forced to duplicate the Hubble sequence.

We thank Greg Wirth for several stimulating conversations on computing
asymmetry in a non-biased way. Both CJC and MAB thank Richard G. Kron
for his ideas on galaxy morphology, the Hubble sequence, and for his
encouragement over the years. We thank the referee, Roberto Abraham
for his thorough reading of the manuscript, and for useful comments
that improved the presentation of this paper. This research was
supported by NASA LTSA grant NAG5-6043, STScI grants AR-7518 and
GO-7875, and research funds from the UW graduate school (MAB); Space
Telescope Science Institute (STScI) is operated by Association of
Universities for Research in Astronomy, Inc., under NASA contract
NAS5-26555. CJC acknowledges the hospitality of Mark Dickinson and
STScI where part of this work was completed.

\clearpage

\begin{figure}
\plotfiddle{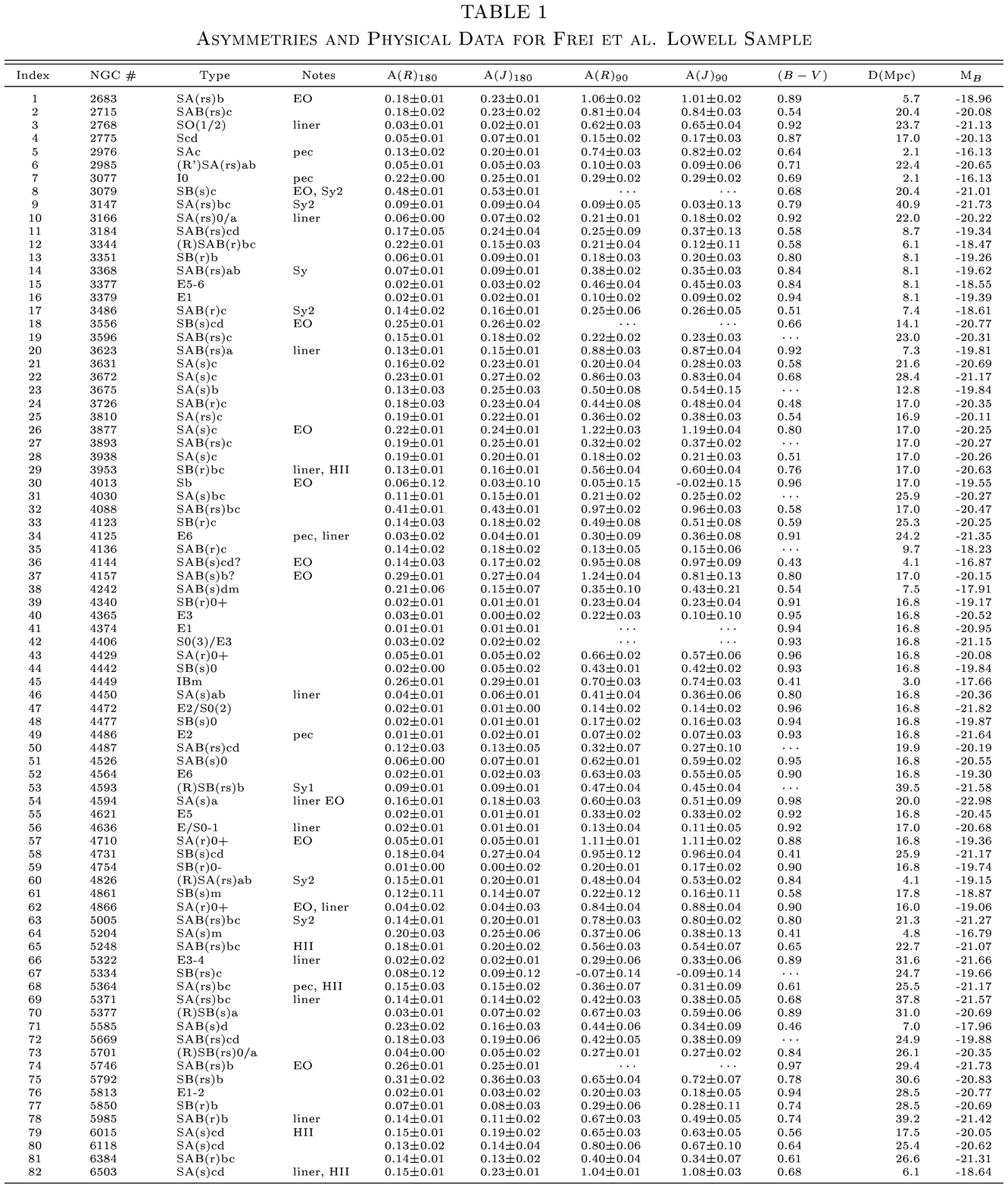}{6.0in}{0}{100}{100}{-310}{-170}
\end{figure}

\clearpage

\begin{figure}
\plotfiddle{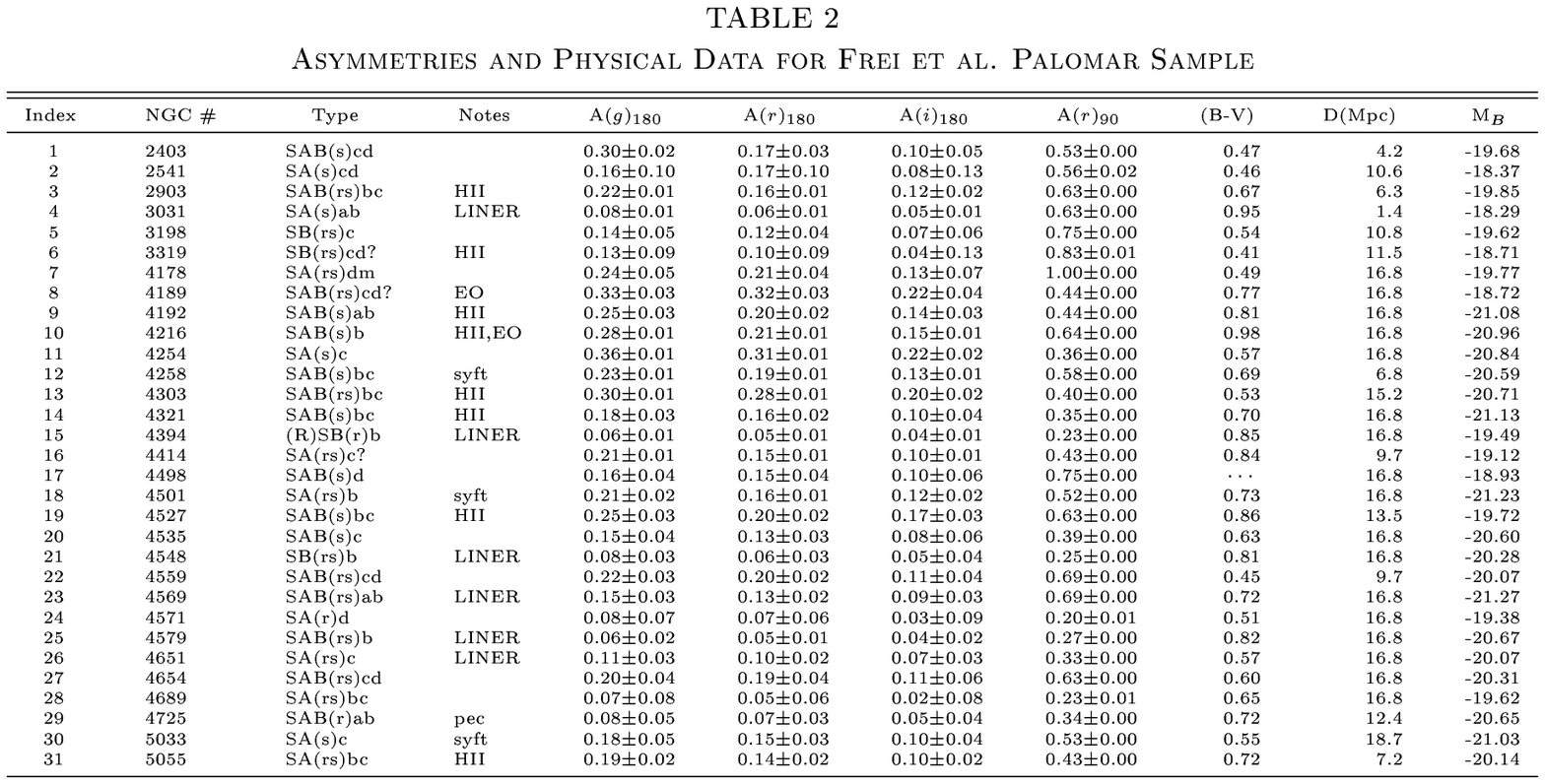}{6.0in}{0}{100}{100}{-310}{-170}
\end{figure}

\clearpage

\begin{figure}
\plotfiddle{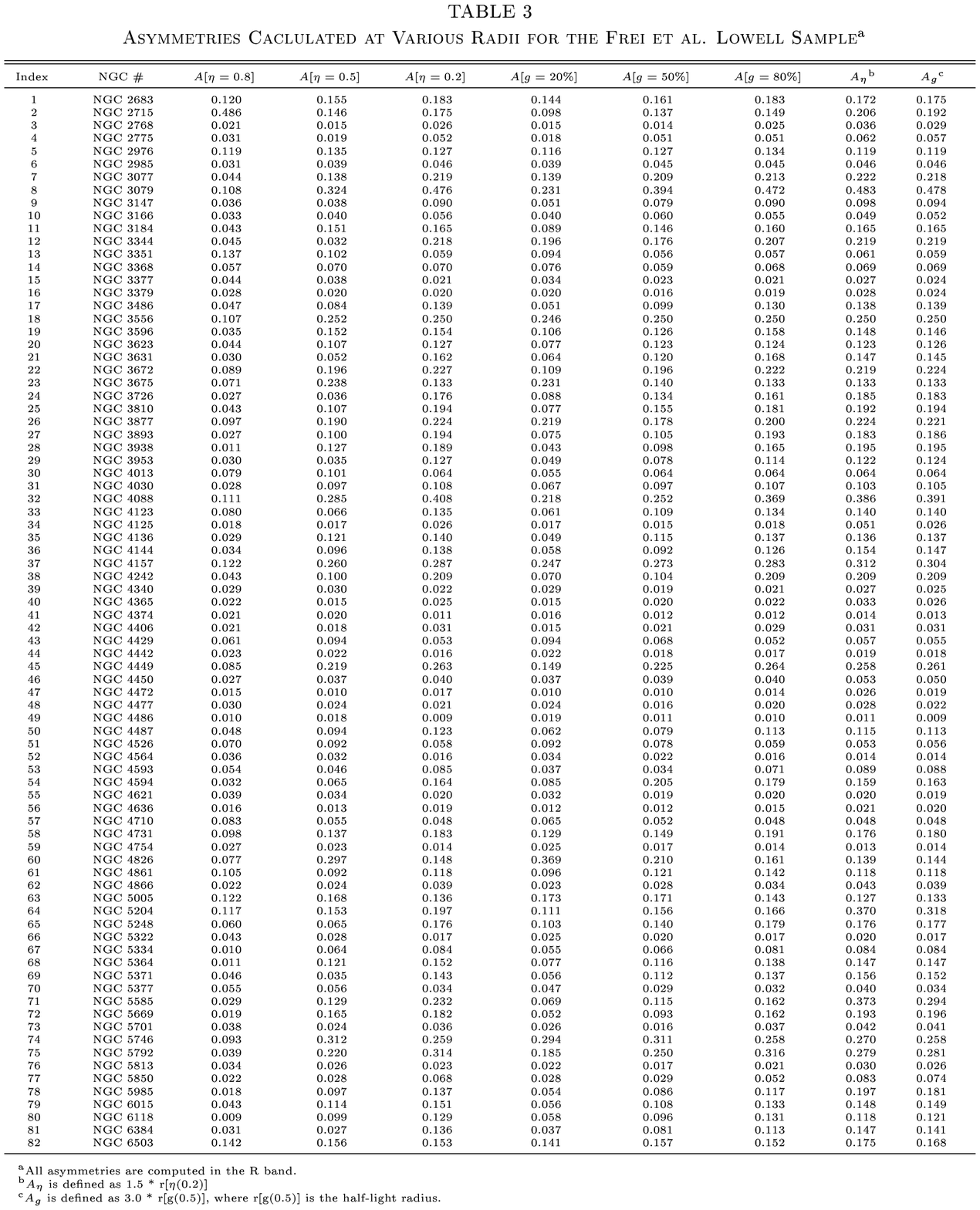}{6.0in}{0}{100}{100}{-310}{-170}
\end{figure}

\clearpage

\begin{figure}
\plotfiddle{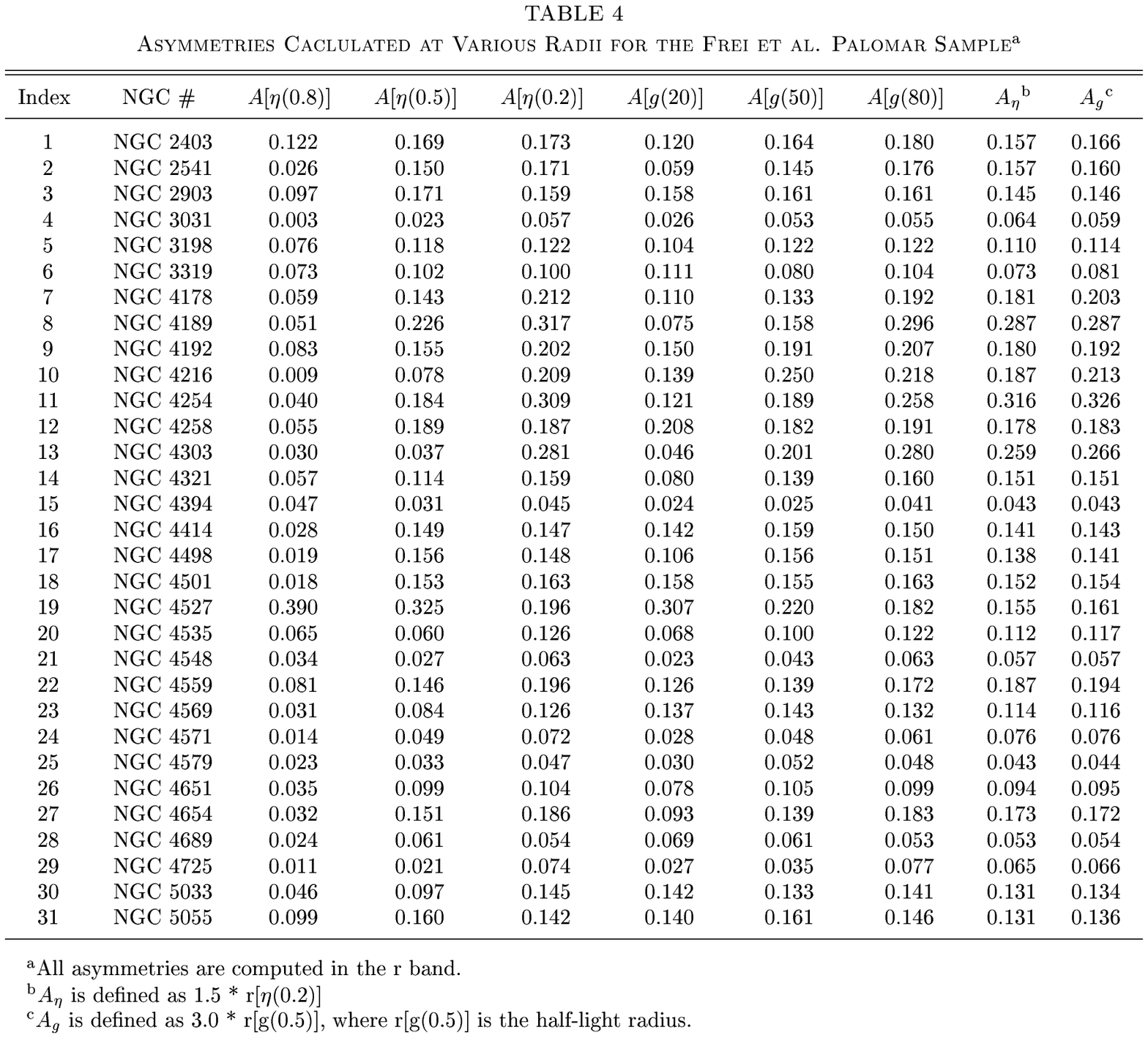}{6.0in}{0}{100}{100}{-310}{-170}
\end{figure}

\clearpage

\figcaption[]{Scatter plot of the red 180\deg asymmetries computed by
the RMS method (Conselice 1997) and the ABS method (Abraham \etal 1996a).  
The two methods produce similar asymmetry values, with a noticeable trend
towards higher asymmetries for more asymmetric galaxies. The open
symbols are Palomar sample galaxies (Gunn $r$ band, Table 2); filled
symbols are the Lowell sample galaxies ($R$ band, Table 1).}

\figcaption[]{Comparison between the 180\deg RMS asymmetry parameter
computed in the $R$ and $B$ bands (filled symbols) and $g$ and $r$
bands (open symbols). The bluer band asymmetries tends to always be
more asymmetric than the redder band asymmetries, and this difference
increases at higher asymmetries. This indicates that young blue stars
(and hence recent star formation) produce asymmetries. There is no
differences in trend or scatter between the $B$ versus $R$ and $g$
versus $r$ data.}

\figcaption[]{As Figure 2, but with the asymmetries computed with the
ABS method. The relationship between the asymmetries does not show
the prominent rise in blue asymmetries as the RMS asymmetry values.}

\figcaption[]{Six representative `Early'-type galaxies from the et
al. (1996) sample. From left to right: (1) the $B_J$ or $g$ band
image. The NGC number, Hubble type, A$_{180}(R)$, and $B-V$ color are
listed in clockwise order starting at the bottom left corner of each
image. (2) Asymmetry, A$_{180}(R)$, as a function of center position,
A(x,y). The grey-scale stretch is from 0.02 (white) to 0.44 (black) in
every panel. The maximum asymmetry value is printed in the upper-right
corner of each panel. (3) A$_{180}(R)$ / A$_{\rm max}$, the asymmetry
as a function of radius (more precisely, the square extraction box
half-width) normalized by the half-light radius, for rotation angle
$\phi = 180$, normalized by the maximum value labeled in
each panel.  The radius where $\eta
= 0.2$ is marked with a dotted vertical line.
 (4) A$_{\phi}(R)$ / A$_{\rm max}$, the
asymmetry as a function of rotation angle ($-180 < \phi < 180$) within
a radius corresponding to $\eta = 0.2$ (see text), normalized by the
maximum value. A$_{\phi}(R)$ is folded every 90 degrees to show the
expected inherent symmetry in an axisymmetric system. Solid lines are
for $0 < \phi < 180$ and dashed lines are for $-180 < \phi < 0$. The
value of A$_{90}(R)$ is labeled.}

\figcaption[]{The same as Figure 4, except for 6 representative
`Intermediate'-type galaxies.}

\figcaption[]{The same as Figure 4, except for 6 representative
`Late'-type
galaxies.}

\figcaption[]{The same as Figure 4, except for an irregular,  disturbed
and edge-on galaxy.}

\figcaption[]{A representation of how asymmetry increases at lower
values of S/N without noise corrections.  Represented here are 35
galaxies from our sample which includes all inclinations and Hubble
types from the entire sample.  The vertical axis $\Delta$A(R) is the
difference between the computed asymmetry at a given S/N (x-axis) and
the original asymmetry value. The S/N is computed within the
$\eta=0.2$ radius. The dashed lines at $\Delta$A($R$) = $\pm$0.1 and the
vertical line at S/N of 100 are for reference. This figure shows that
without correcting for the S/N, the asymmetry of a galaxy becomes
large even at moderately large S/N values.}

\figcaption[]{Plot of the asymmetry within the $\eta = 0.2$ radii as a
function of the S/N as in Figure 8, except here noise
corrected. Reliable values, i.e. within 0.05 of the original value, of
the asymmetry can be computed with the algorithm presented in this
paper down to a S/N of about 100.  Dashed lines at $\Delta$A($R$) =
$\pm$0.1 and the vertical line at S/N of 100 are for reference.}

\figcaption[]{Color-Asymmetry diagram.  This physical-morphological
diagram (as defined in text) can be used to determine the population
of galaxies in a large sample. Labeled galaxies are undergoing
interactions or galaxy mergers.  Three features stand out in this
diagram: (1) The early-type disks/spheriods are well separated from
the late type disks. (2) The interacting galaxies deviate from the
color-asymmetry sequence such that they are too asymmetric for their
color.  Their asymmetries are not caused solely by star formation, but
in part result from tides distorting their structure. (3) The edge-on
galaxies also deviate, and generally are too asymmetric or red, both
effects caused by dust lanes affecting the morphology (higher
asymmetry) and the color (redder).}

\figcaption[]{The relationship between A$_{90}$ and A$_{180}$.
Plotted here as different symbols are spheriods, early type disks,
late type disks, irregulars, and edge-on galaxies.  For a given range
of A$_{180}$ there is a high scatter in the A$_{90}$ values.  The
edge-on galaxies stand out in this diagram as being the systems with
the highest A$_{90}$ values for their A$_{180}$ values.  There is also
a fairly sharp division between spheriods/early-type disks and
late-type disks.  We also find no spheriods and few early type disks
with A$_{90}$ $>$ 0.7.  This type of plot can therefore be used to
determine general properties of a sample of galaxies.}

\figcaption[]{Plot of the asymmetries in the $R$ and $B$ bands computed
by rotating the galaxy by 90\deg instead of the nominal 180\deg.  The
mean difference A($B$)$_{90}$ - A($R$)$_{90}$ is near zero for all
A($R$)$_{90}$, in contrast to A($B$)$_{180}$ - A($R$)$_{180}$, which
systematically deviates to positive values for larger
A($R$)$_{180}$. Since most galaxies are not perfectly round, A$_{90}$
measures the ellipticity, or deviations in shape from a perfect
circle. The edge-on galaxies always have the highest values of
A$_{90}$. The consistency between the two different wavelengths
indicates that contributions to A$_{90}$ are mostly due to global
shape. However, higher values are A($B$), compared with A($R$), are
seen but only at the largest values of A$_{180}$, indicating that star
formation is not a significant addition to A$_{90}$, but does
contribute somewhat as would be expected.}

\figcaption[]{Plot of the axis ratio for our sample (log(b/a) from
RC3) versus the asymmetry computed with a rotation angle of 90\deg.
This strong correlation is evidence that A$_{90}$ is primarily a
measure of the azimuthal {\em shape} of a galaxy, rather than physical
effects, such as star formation, or dynamical effects which primary
cause A$_{180}$ asymmetries.  The bottom plot shows A$_{90}$ -
A$_{180}$ plotted instead of A$_{90}$ showing a better correlation,
indicating that there is some flocculent asymmetry contributing to
A$_{90}$. Linear regressions are for illustrative purposes.}

\figcaption[]{The color-asymmetry diagram showing the distribution of
galaxies with high axis rations [b/a].  A principle cause of the
scatter in the normal galaxy-sequence of Figure 10 are reddening and
morphological changes that occur for inclined galaxies.  The galaxies
with the highest inclinations are the ones that deviate the most from
this sequence.}

\figcaption[]{The color corrected color-asymmetry diagram.  This
version of the color-asymmetry diagram uses dereddened $(B-V)$ colors,
corrected for inclination effects (de Vaucouleurs et al. 1991).  The
normal galaxy sequence is tighter, and many of the edge-on
galaxies nearly coincide with other galaxies having similar colors and
asymmetries.  This figure, and Figure 14 shows how this diagram is
being affected by inclination. }

\figcaption[]{Asymmetry-concentration diagram.  This diagram is
similar to the one used by Abraham \etal (1996b) for galaxies in the
Hubble Deep Field.  The spheriods stand out in this diagram as being
the symmetric and most concentrated objects, while the late disks
are less concentrated and more asymmetric.  There is not however, as clean
a division between early and late disks as seen in the color-asymmetry
diagram (Figure 12).  Interestingly, nearly all of the edge-on disks fit 
between the two dashed lines.  The four interacting galaxies still stand out
in this diagram.}

\figcaption[]{The asymmetry distribution as a function of
Hubble T-type.  There is a general increase in average asymmetry at
later types, but there is a large scatter in asymmetry at each Hubble
type, similar to the pattern found for most physical parameters.}

\figcaption[]{(a) Top panel. $B-V$ vs $B$ absolute magnitude, with
different symbols distinguishing morphological types as in Figure 16.  (B)
Bottom panel. The red asymmetry parameter ($R$ or $r$) plotted as a
function of the absolute magnitude. As shown in Conselice (1997) there
is little correlation between the intrinsic brightness and asymmetry
of a galaxy for the narrow range of magnitudes in the Frei \etal
sample.}

\figcaption[]{Asymmetry as a function of spatial resolution.  The
parameter $\epsilon$ is the ratio of the angular diameter
corresponding to 0.5 h$_{75}^{-1}$ kpc divided by the angular
resolution.  From this plot it can be seen that the asymmetry of a
galaxy is measurable to better than 10\% down to $\epsilon$ $\approx$
1.}

\figcaption[]{Apparent size versus redshift for a metric length of
0.5 h$_{75}^{-1}$ kpc and several values of $\Omega$ composed of
non-relativistic matter and a cosmological constant. The solid lines
are for $\Omega_{total} = 0.1$ and 0.3, and $\Omega_\Lambda = 0$,
i.e. q$_0 = 0.05$ and $0.15$ for the bottom and top solid curves,
respectively. The dashed curves are for $\Omega_{total} = 1$, with
$\Omega_\Lambda = 0.9$ and $0.3$, i.e. q$_0 = -0.850$ and $-0.550$ for
the bottom and top dashed curves, respectively. Accessible apparent
sizes for different instruments classes are indicated by the shading:
modern ground-based telescopes at good sites in ambient conditions
(ground/ambient, e.g. WIYN); ground-based telescopes with adaptive
optics (ground/AO, e.g. CFHT); space-based telescopes (HST,
NGST). Since scales of 0.5 h$_{75}^{-1}$ Mpc must be resolved to
accurately measure rotational asymmetries, cosmological measurements
of asymmetry require spatial resolutions of 0.1 arcsec or better.}

\clearpage

\setcounter{figure}{0}

\begin{figure}
\plotfiddle{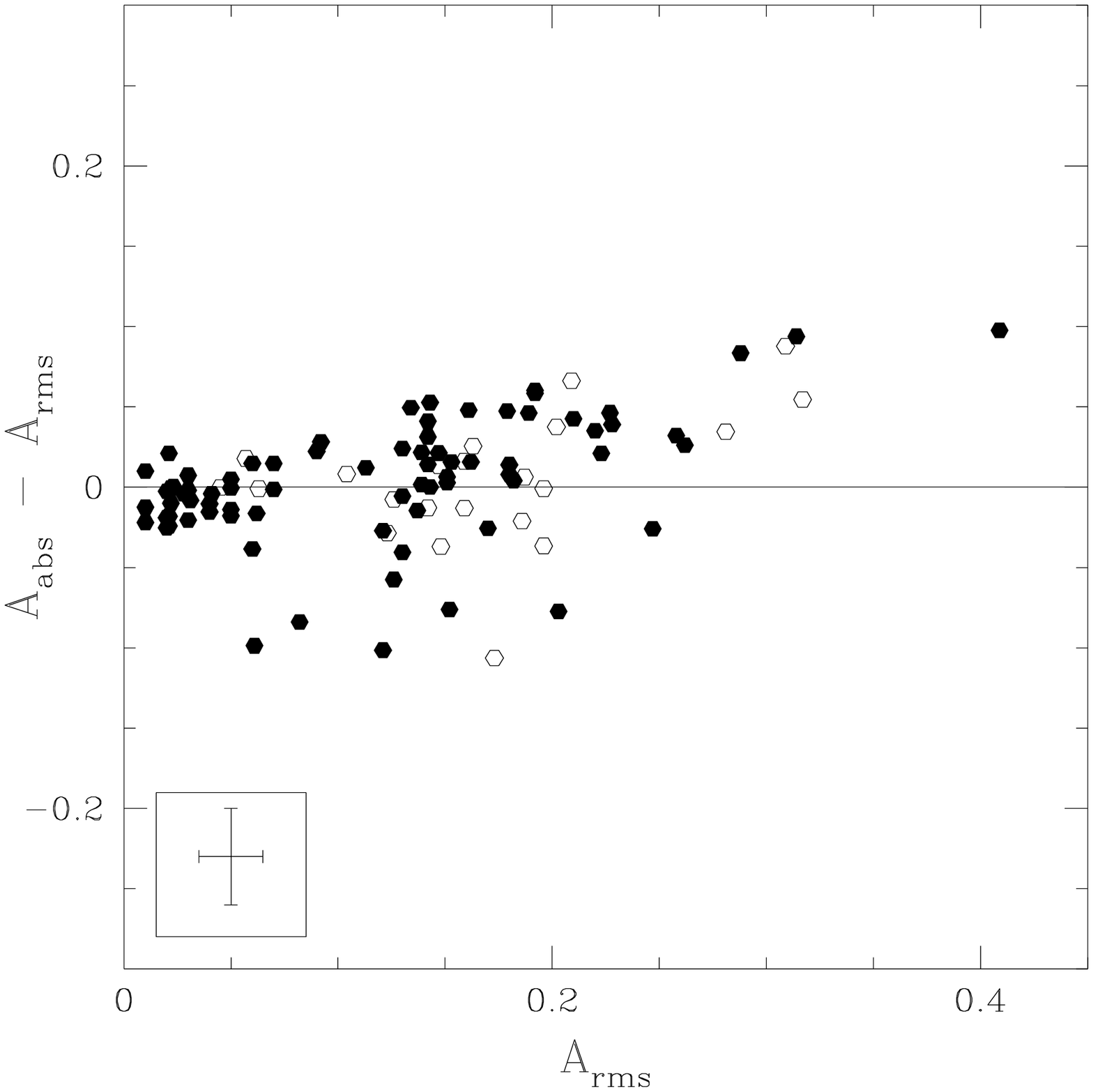}{6.0in}{0}{80}{80}{-250}{-100}
\caption{}
\end{figure}

\clearpage

\begin{figure}
\plotfiddle{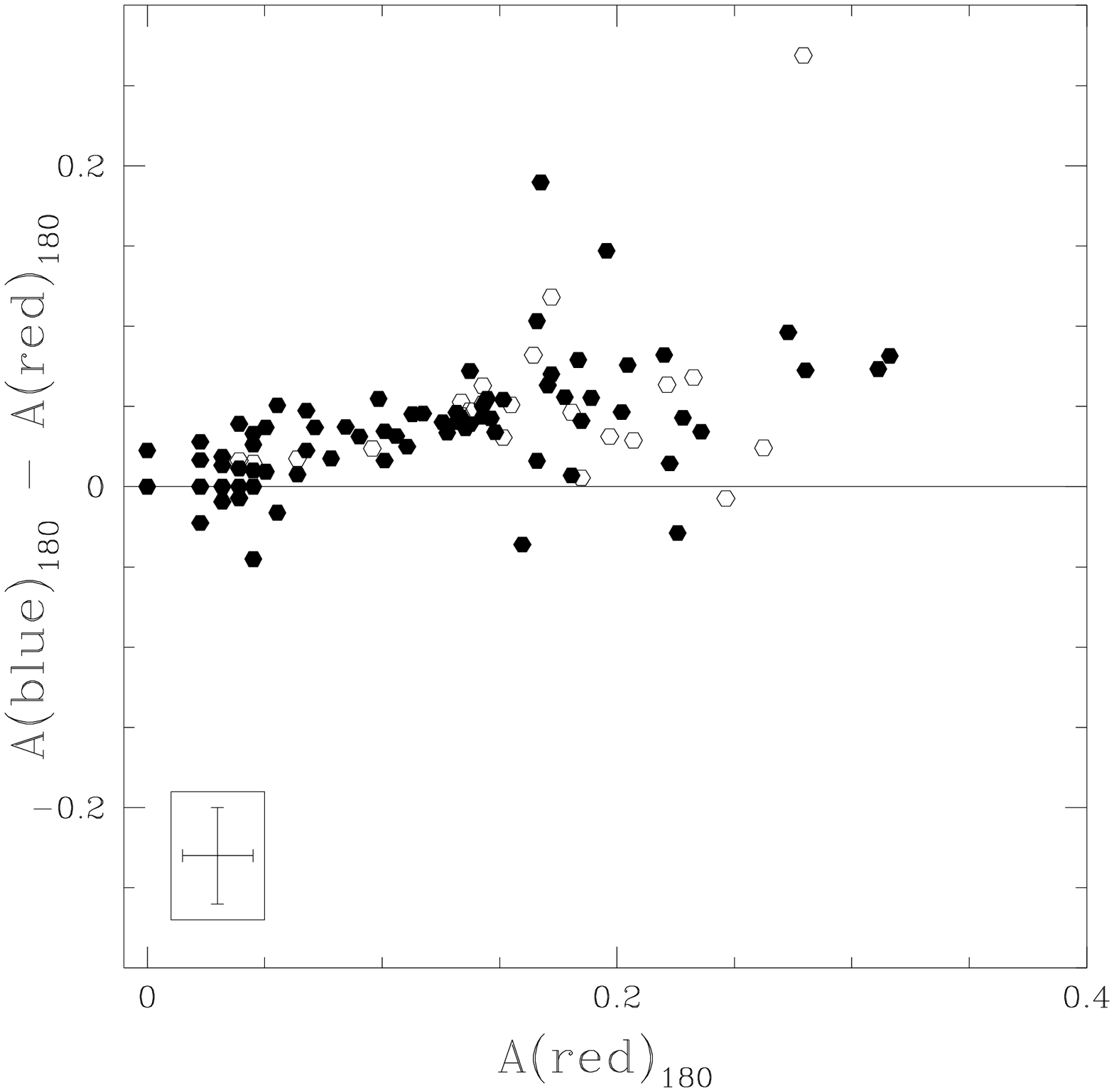}{6.0in}{0}{80}{80}{-250}{-100}
\caption{}
\end{figure}

\clearpage

\begin{figure}

\plotfiddle{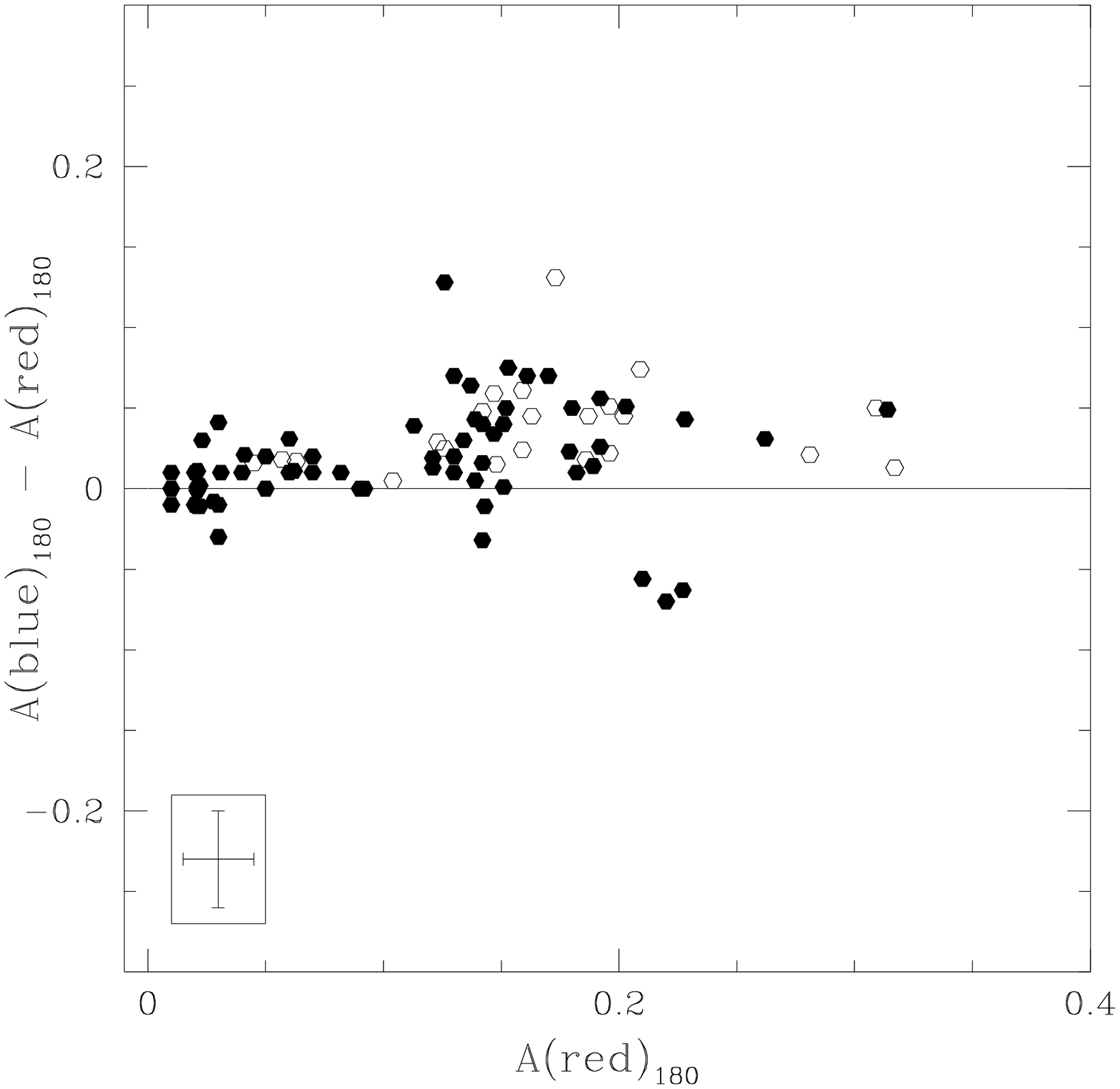}{6.0in}{0}{80}{80}{-250}{-100}
\caption{}
\end{figure}

\clearpage

\begin{figure}
\plotfiddle{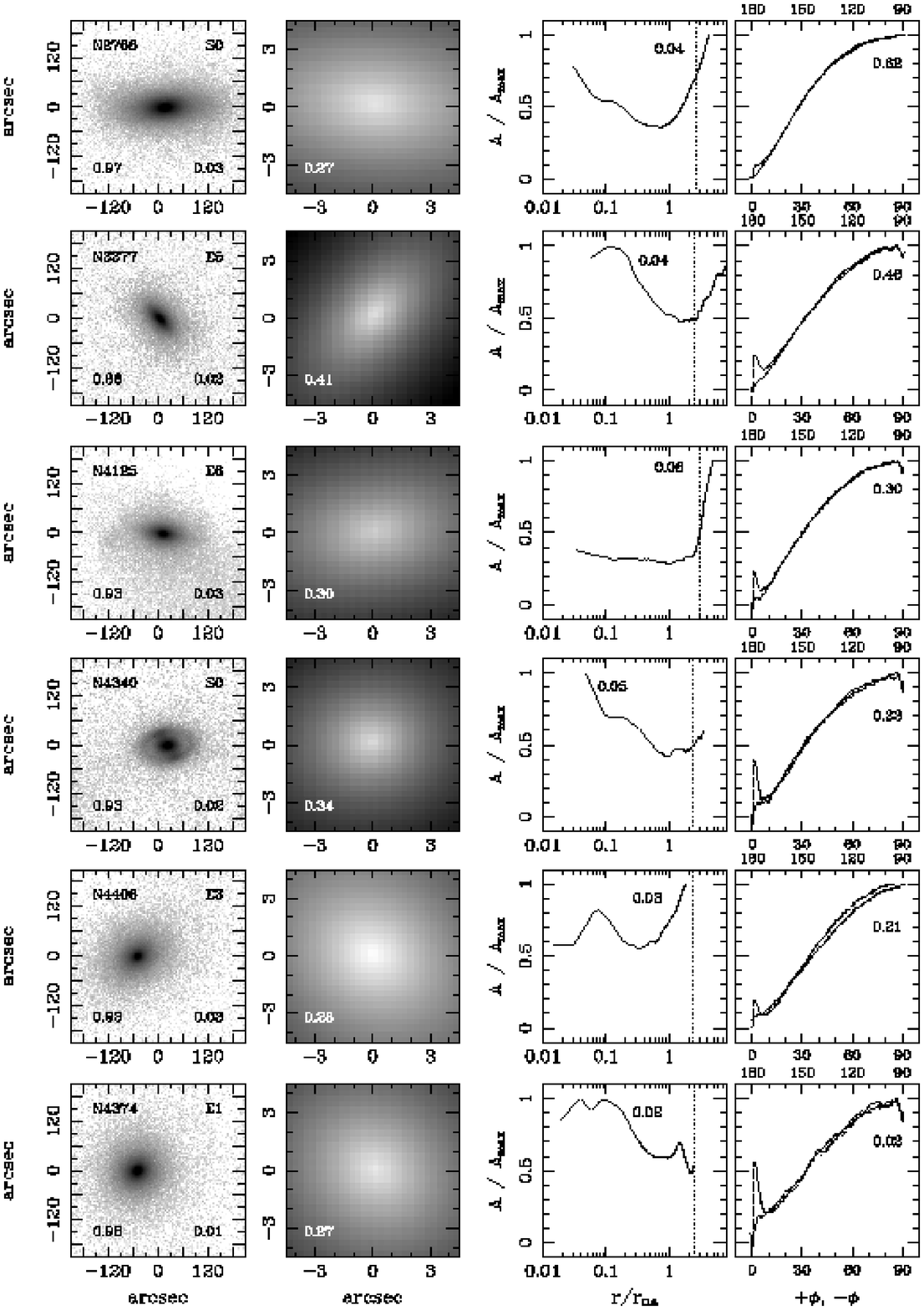}{7.0in}{0}{80}{80}{-250}{-100}
\vskip 1.0in
\caption{}
\end{figure}

\clearpage

\begin{figure}
\plotfiddle{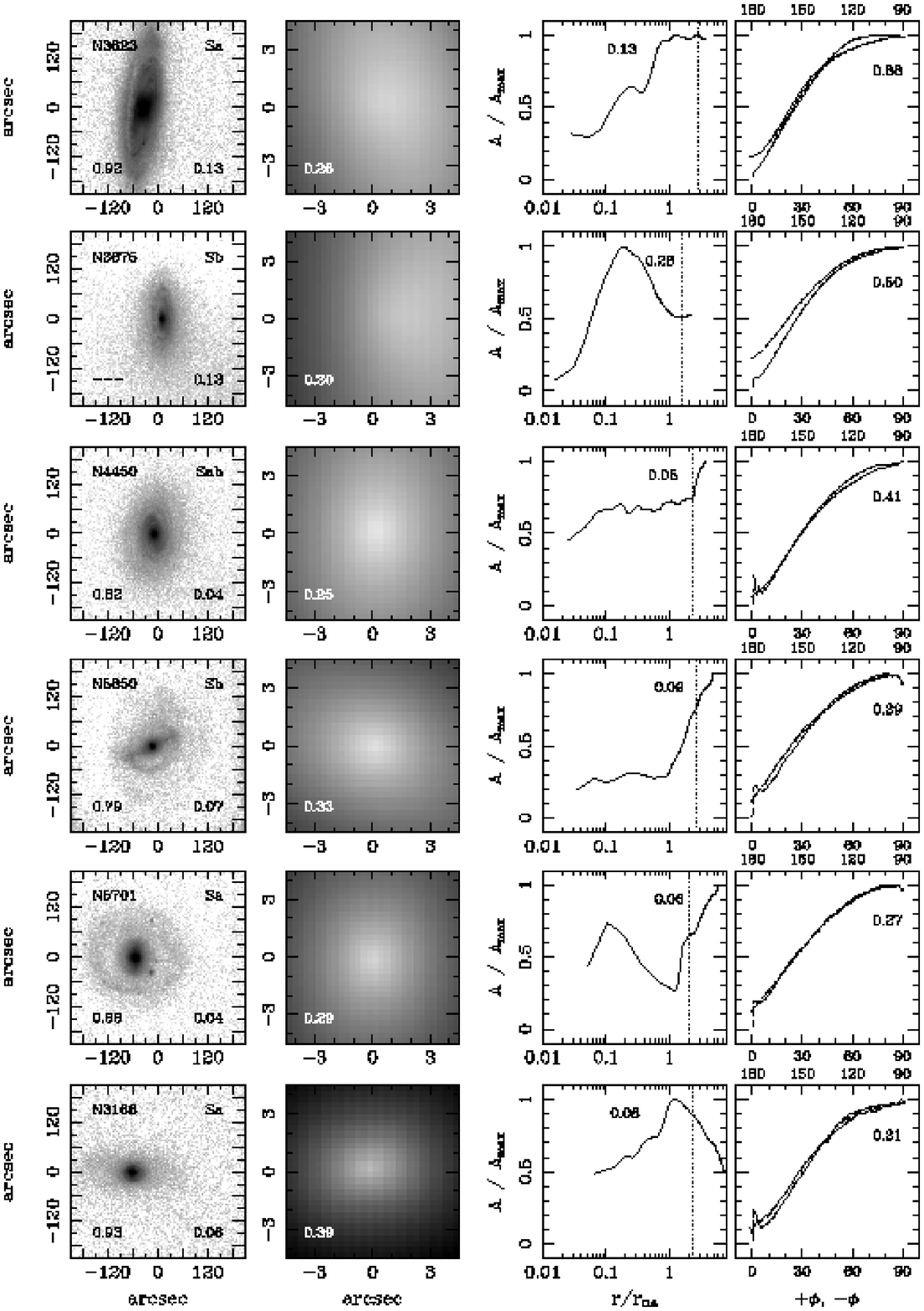}{7.0in}{0}{80}{80}{-250}{-100}
\vskip 1in
\caption{}
\end{figure}

\clearpage

\begin{figure}
\plotfiddle{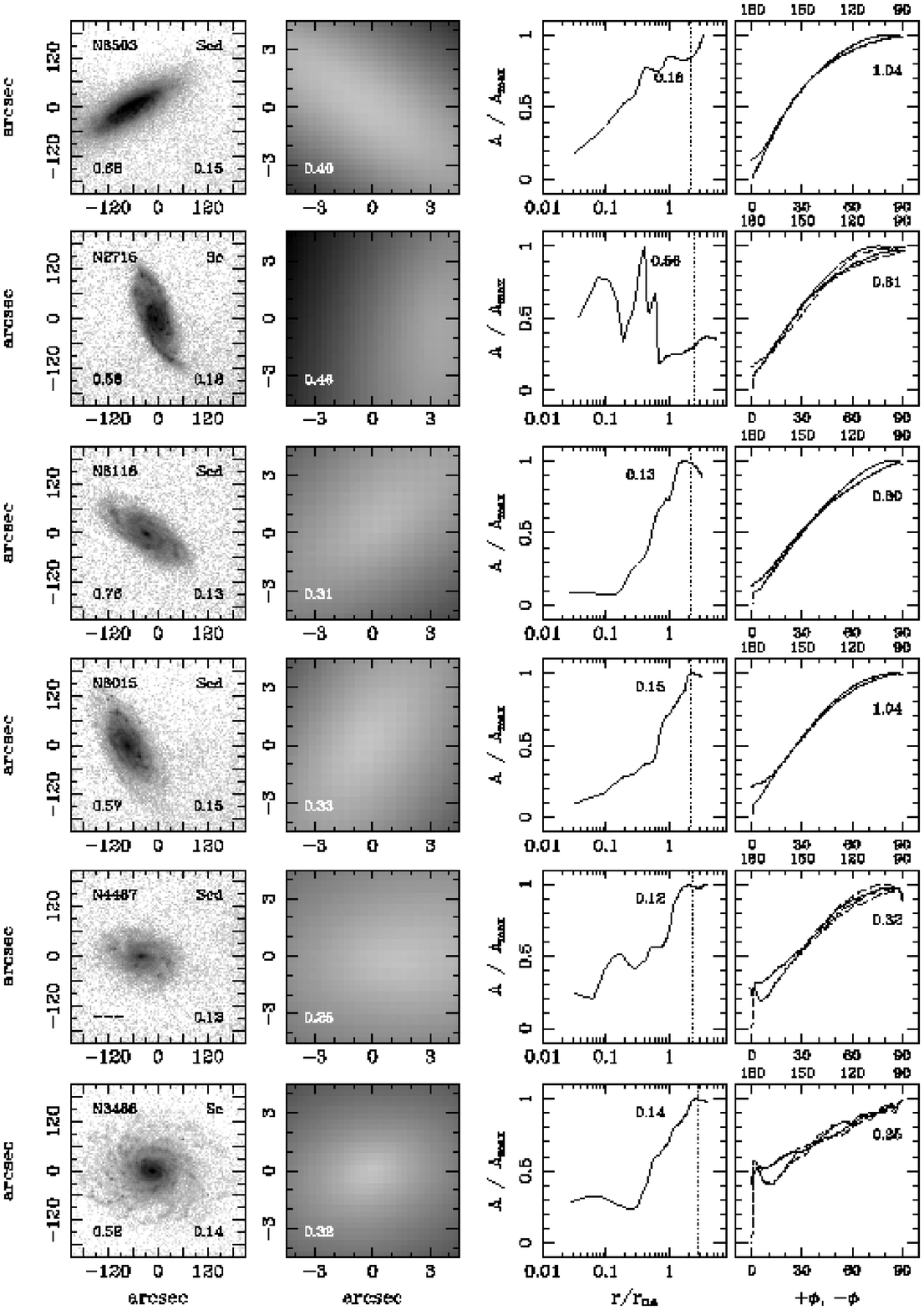}{7.0in}{0}{80}{80}{-250}{-100}
\vskip 1in
\caption{}
\end{figure}

\clearpage

\begin{figure}
\plotfiddle{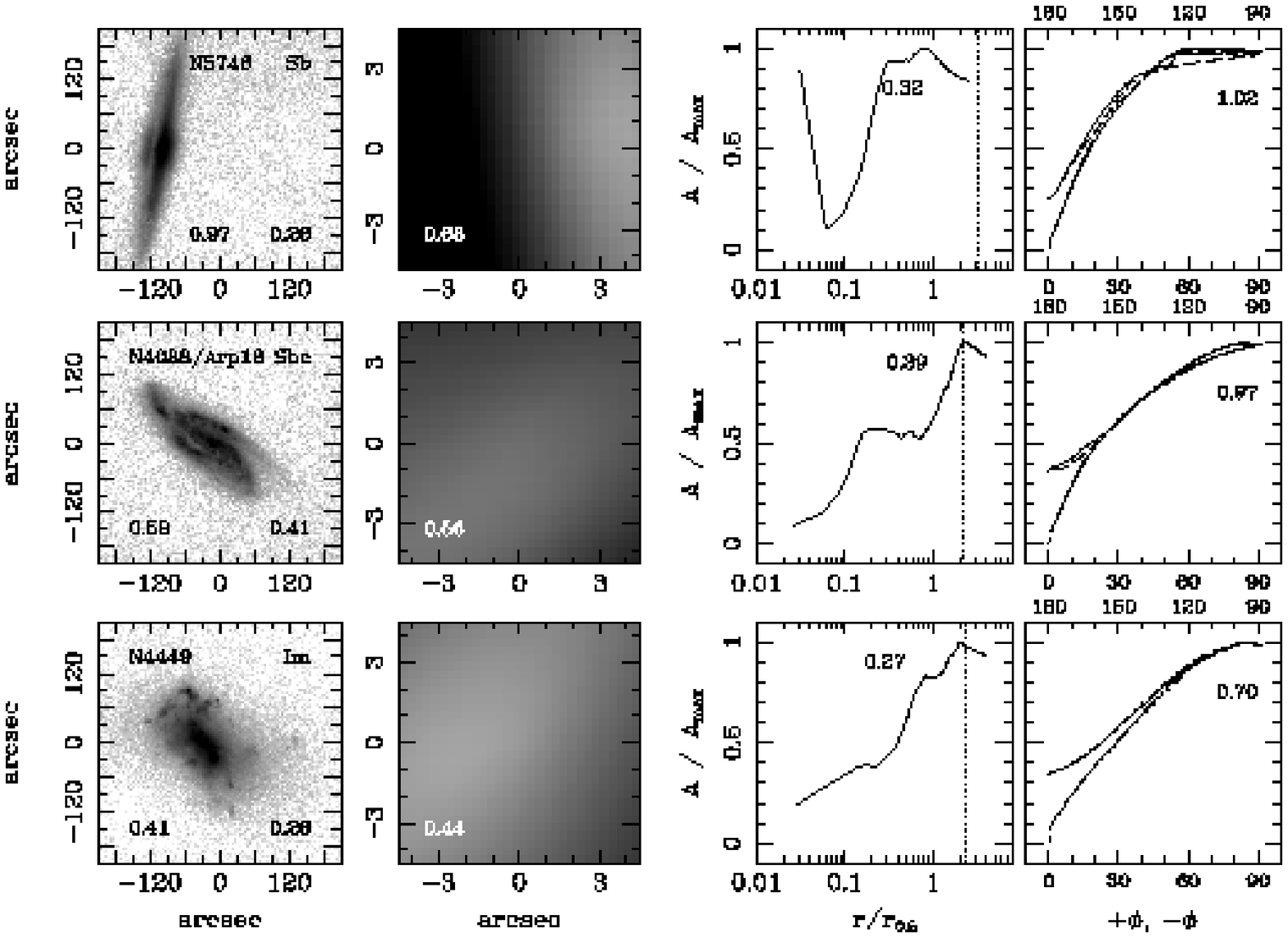}{4.5in}{-90}{60}{60}{-255}{400}
\caption{}
\end{figure}

\clearpage

\begin{figure}
\plotfiddle{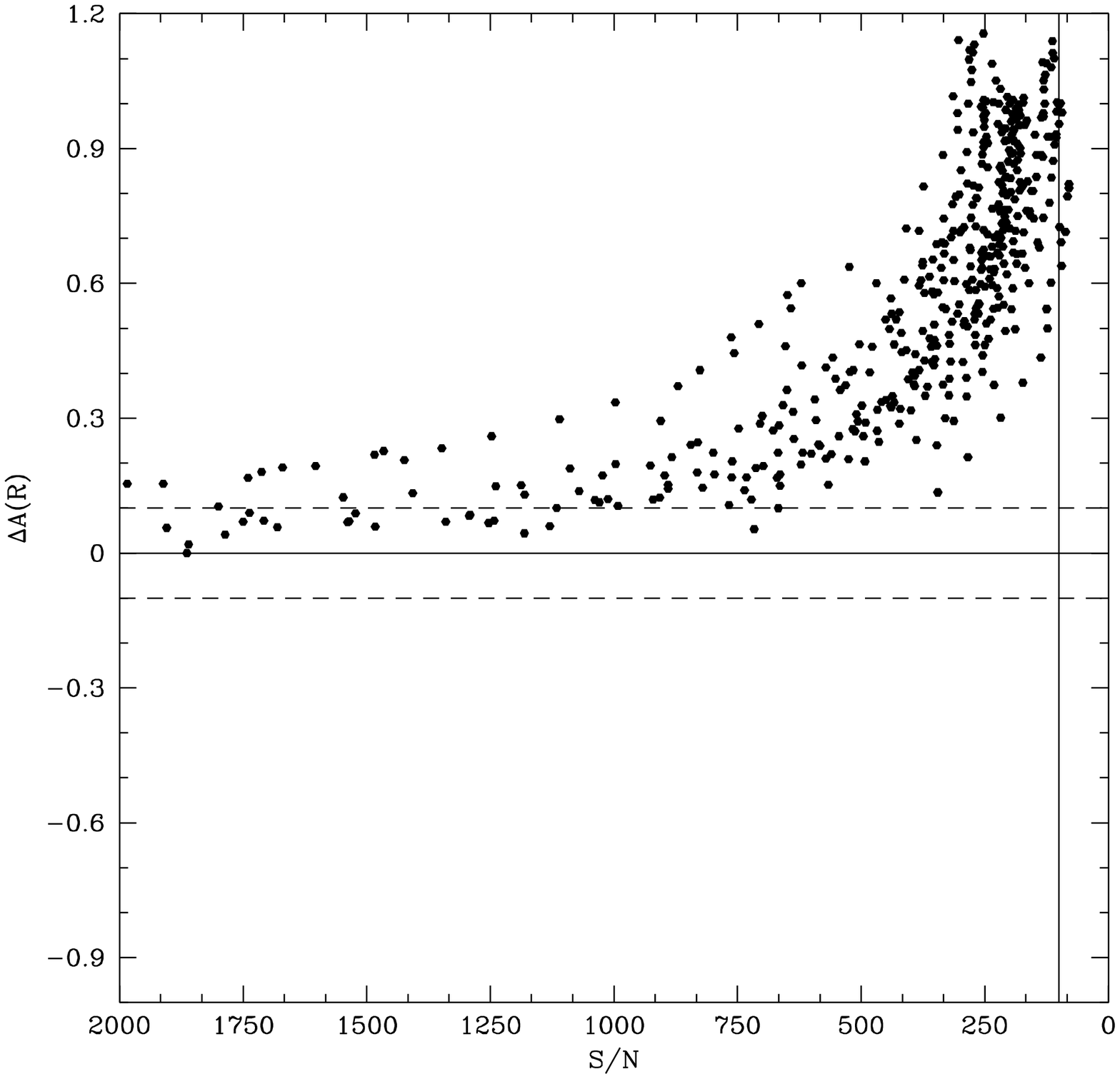}{6.0in}{0}{80}{80}{-250}{-100}

\caption{}
\end{figure}

\clearpage

\begin{figure}
\plotfiddle{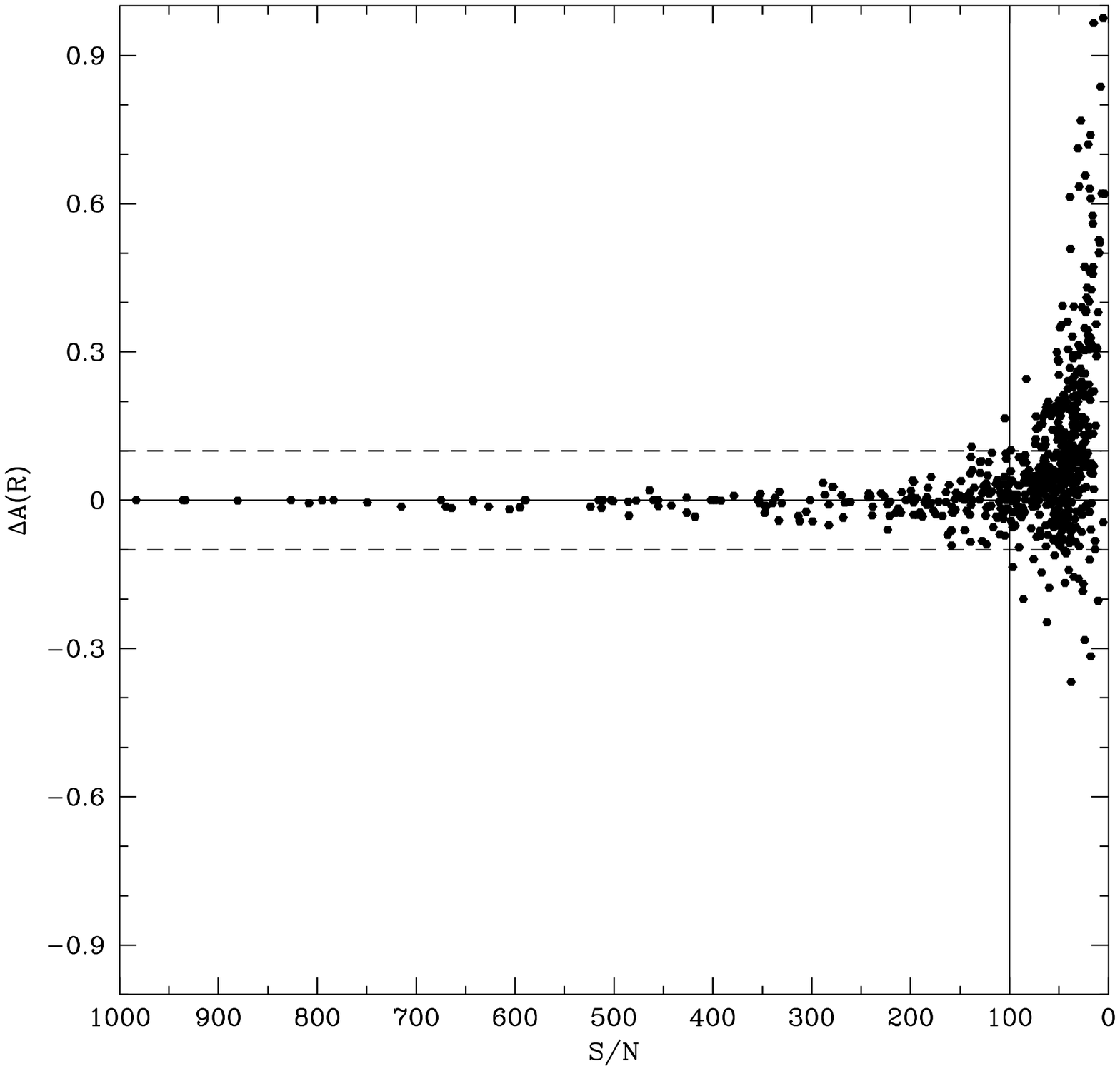}{6.0in}{0}{80}{80}{-250}{-100}
\caption{}
\end{figure}

\clearpage

\begin{figure}
\plotfiddle{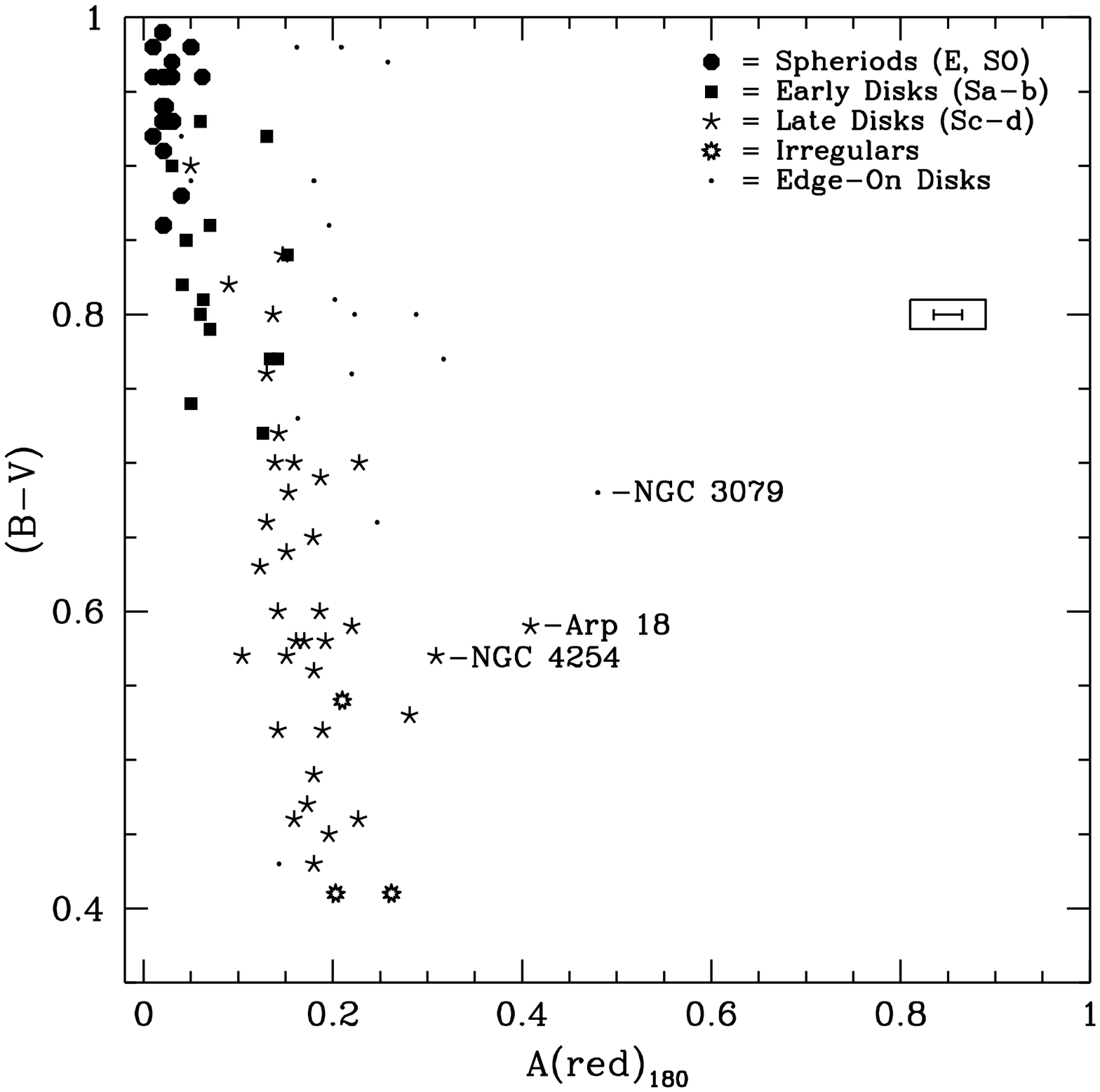}{6.0in}{0}{80}{80}{-250}{-100}
\caption{}
\end{figure}

\clearpage

\begin{figure}
\plotfiddle{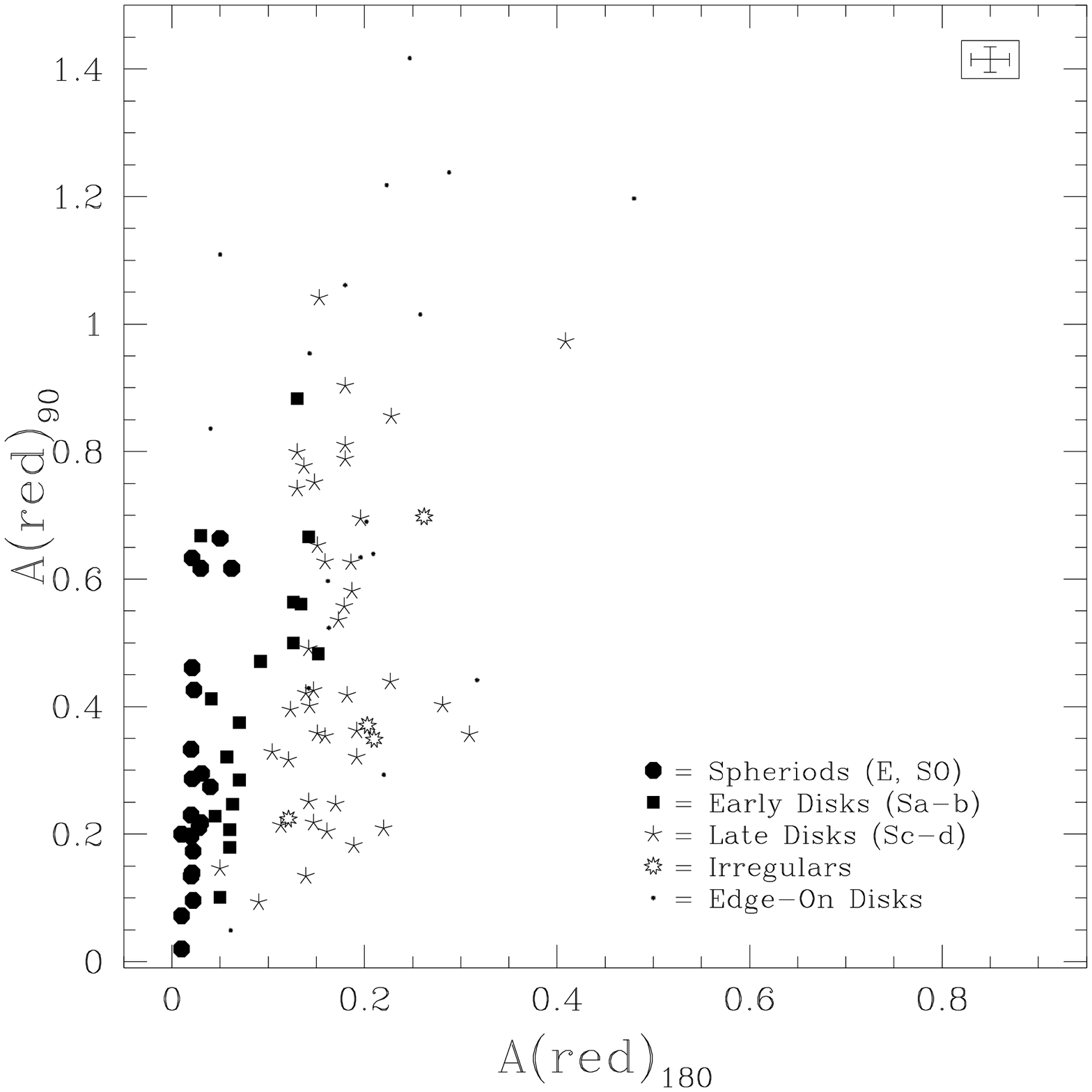}{6.0in}{0}{80}{80}{-250}{-100}
\caption{}
\end{figure}

\clearpage

\begin{figure}
\plotfiddle{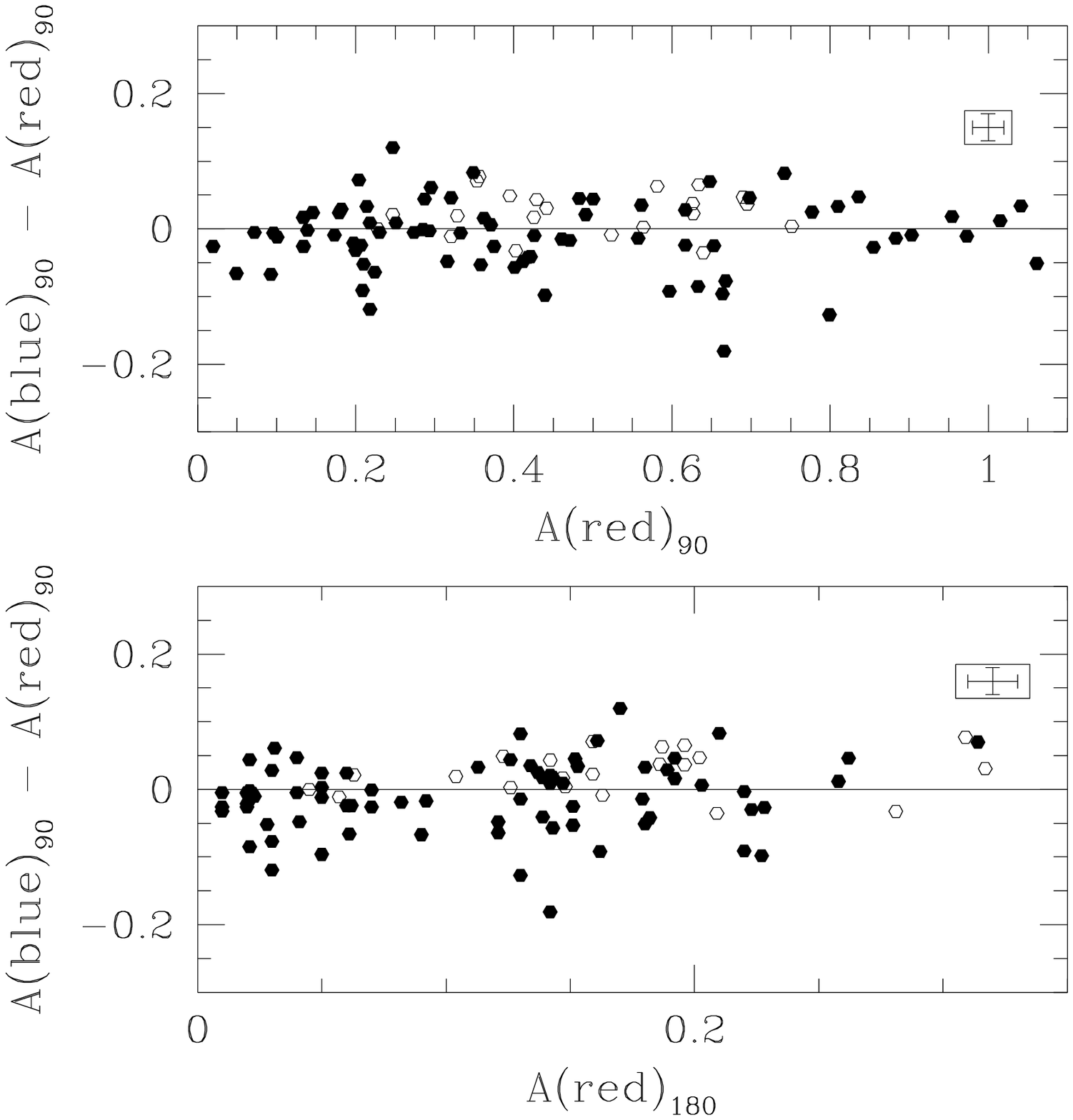}{6.0in}{0}{80}{80}{-250}{-100}
\caption{}
\end{figure}

\clearpage

\begin{figure}
\plotfiddle{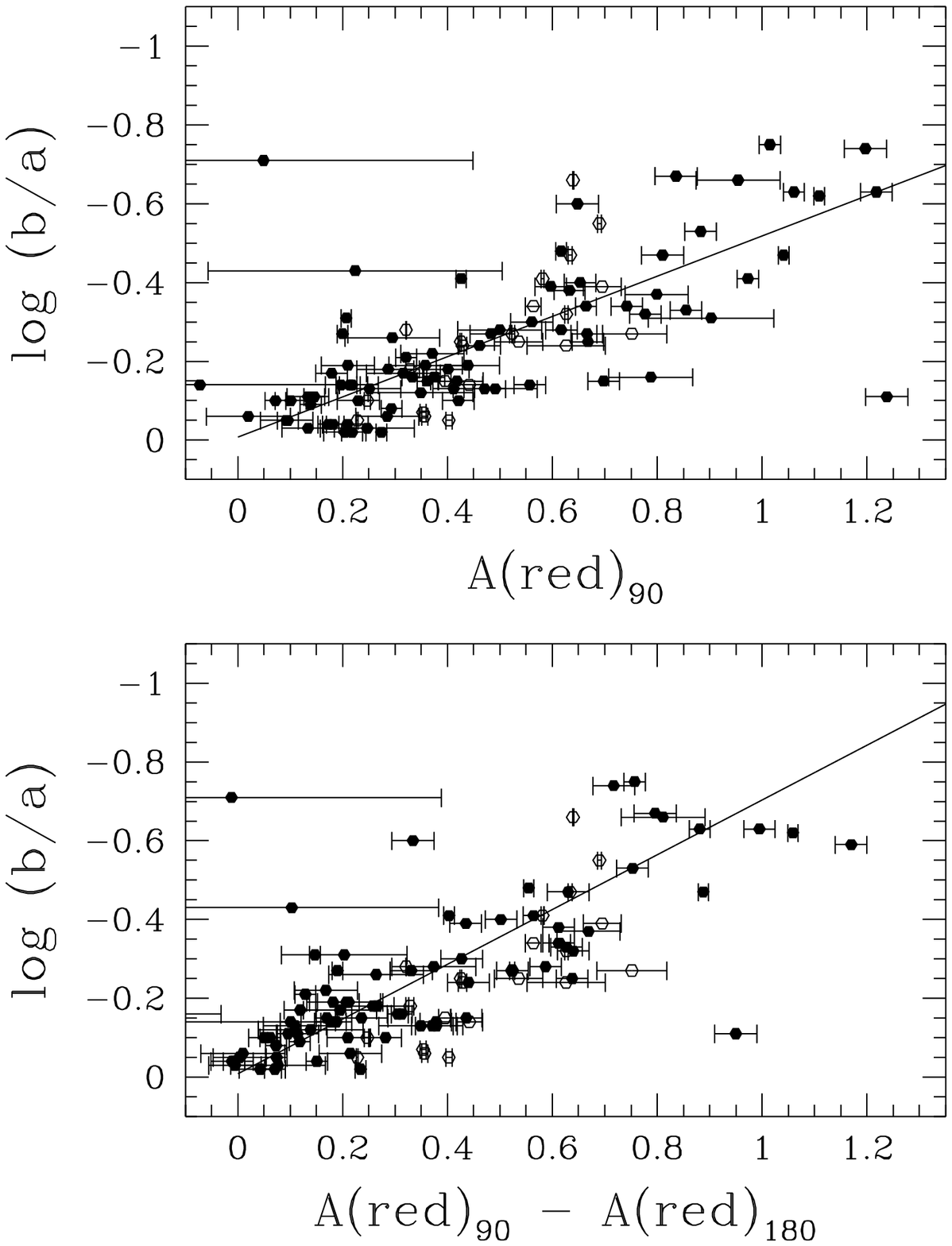}{6.0in}{0}{80}{80}{-250}{-100}
\caption{}
\end{figure}

\clearpage

\begin{figure}
\plotfiddle{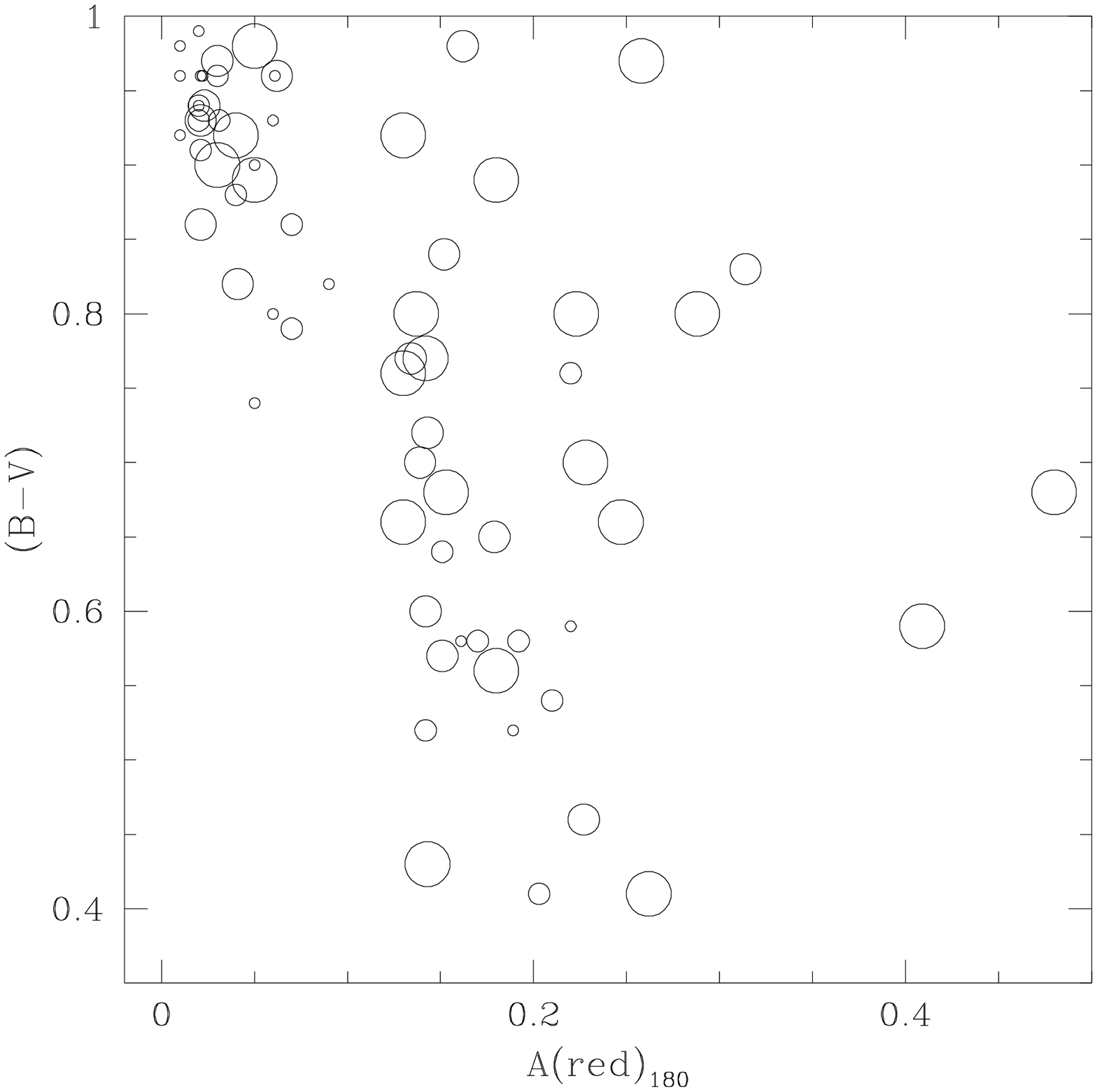}{6.0in}{0}{80}{80}{-250}{-100}
\caption{}
\end{figure}

\clearpage

\begin{figure}
\plotfiddle{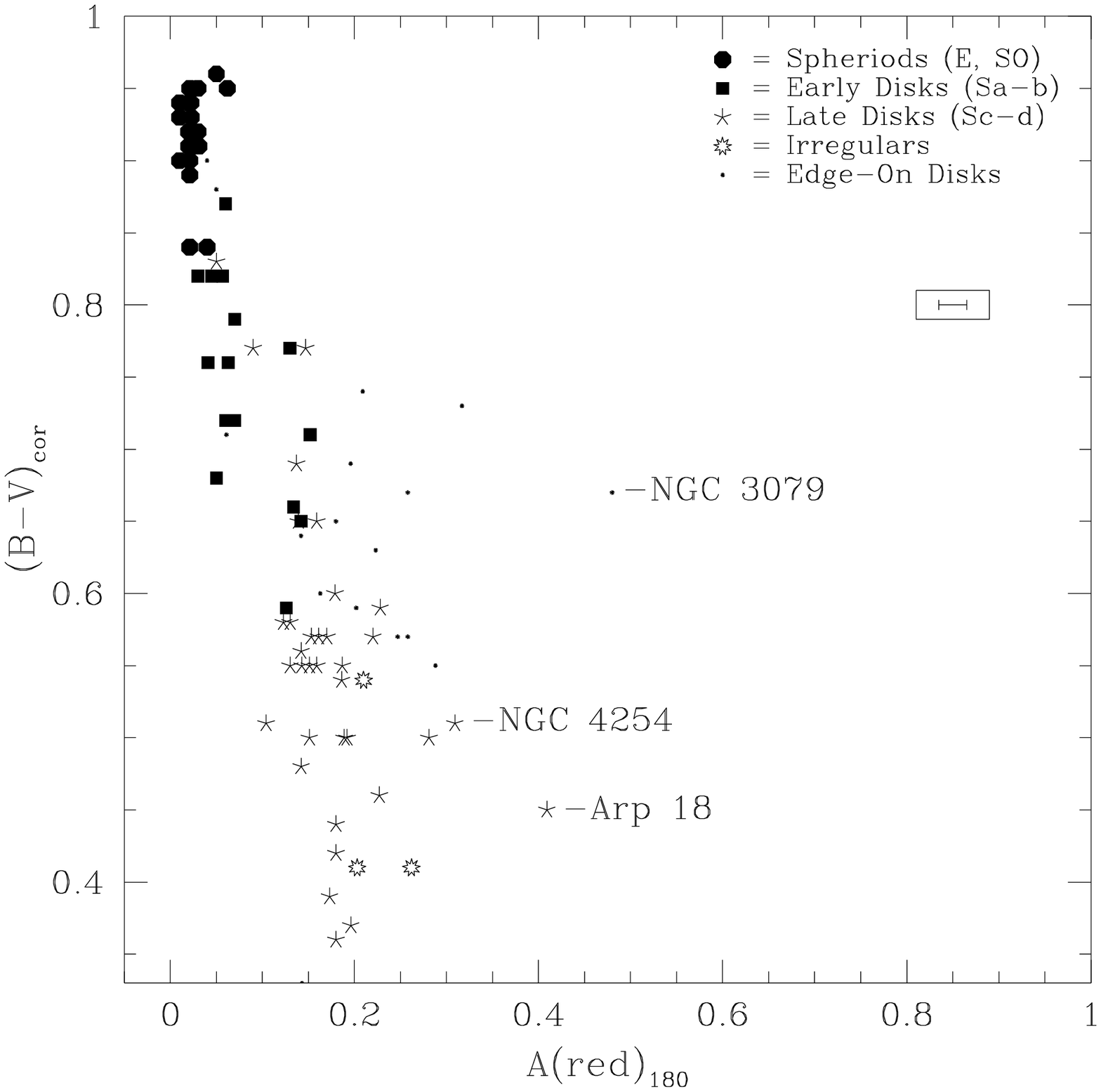}{6.0in}{0}{80}{80}{-250}{-100}
\caption{}
\end{figure}

\clearpage

\begin{figure}
\plotfiddle{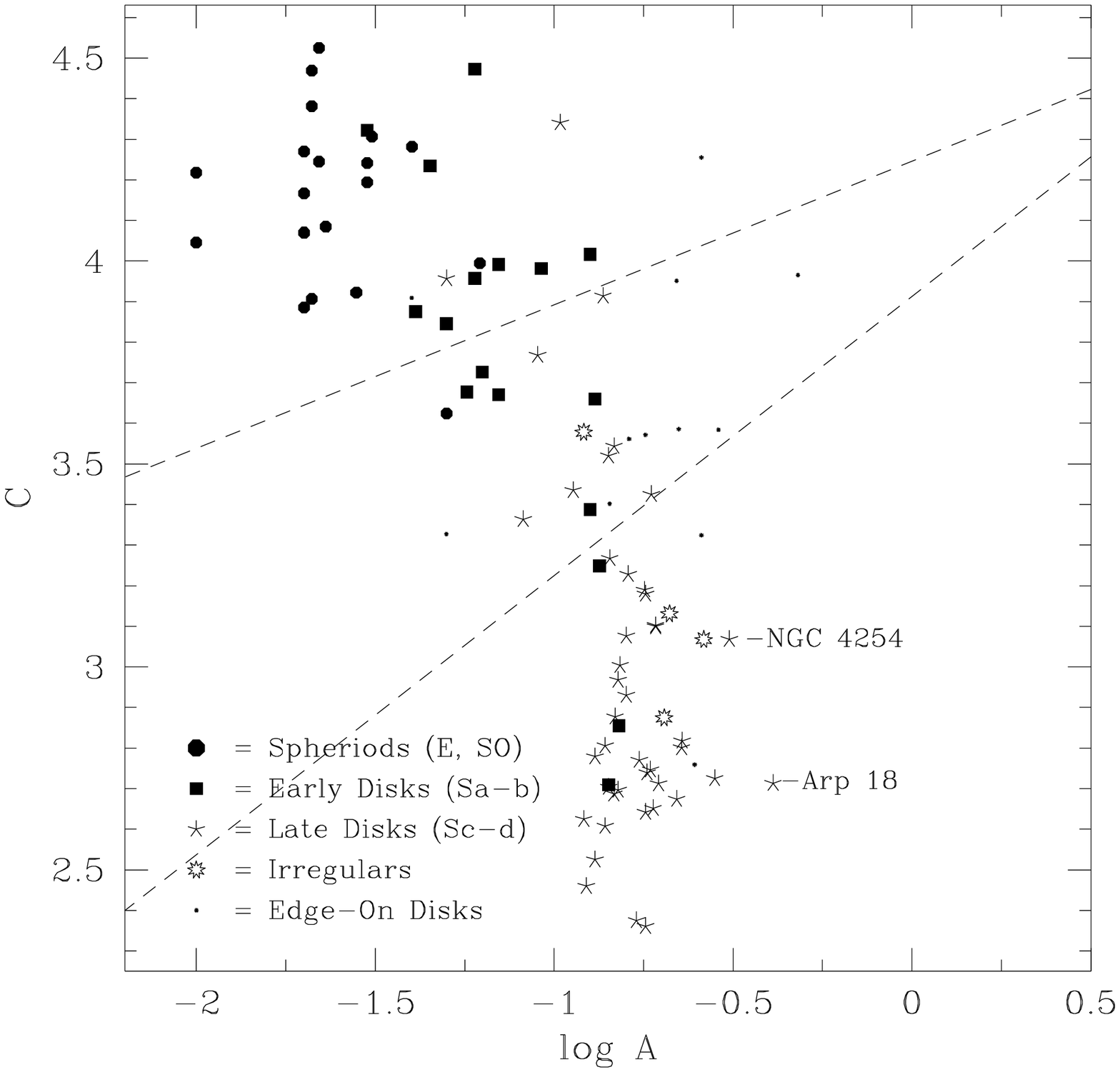}{6.0in}{0}{80}{80}{-250}{-100}
\caption{}
\end{figure}

\clearpage

\begin{figure}
\plotfiddle{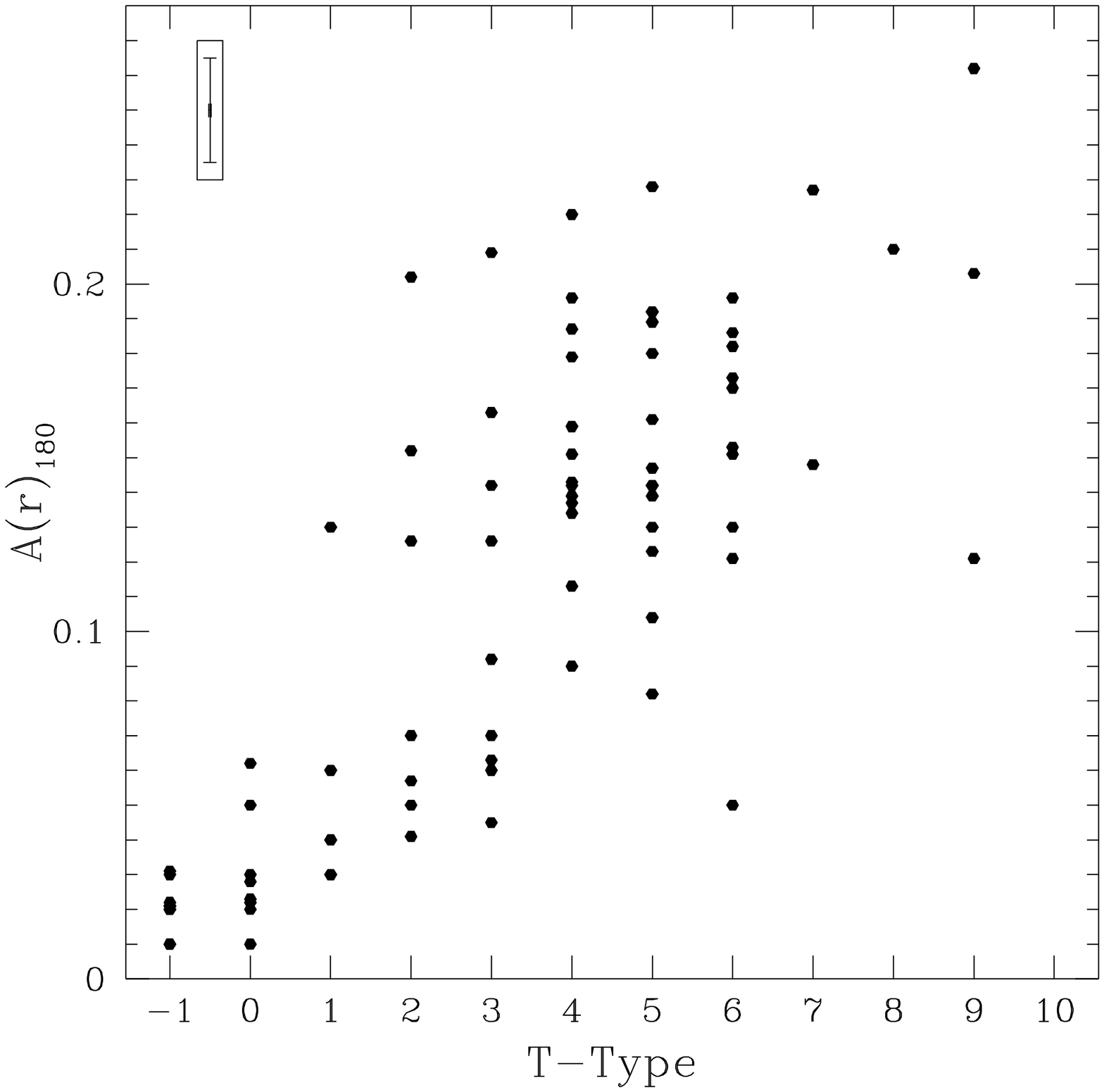}{6.0in}{0}{80}{80}{-250}{-100}
\caption{}
\end{figure}

\clearpage

\begin{figure}
\plotfiddle{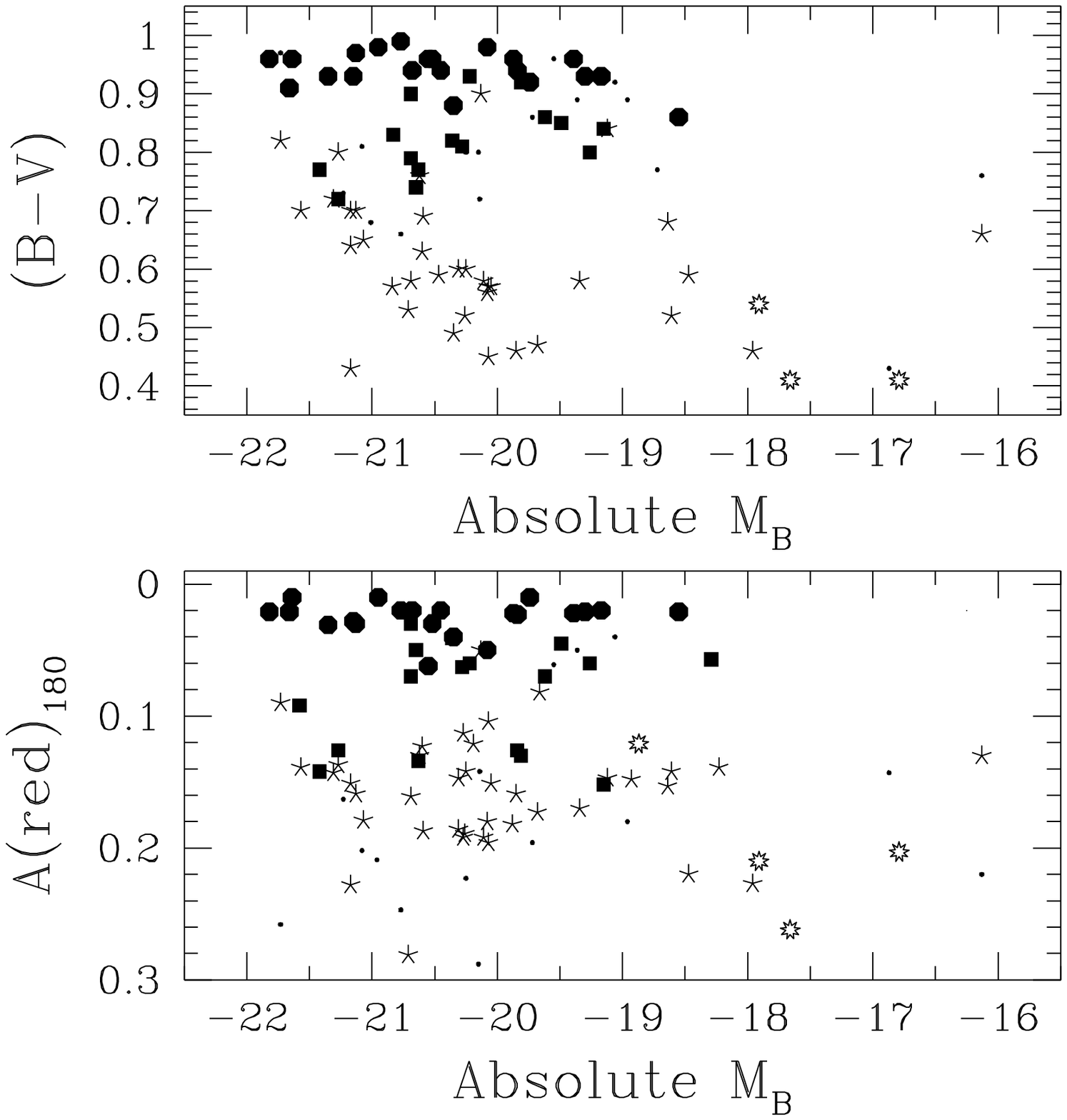}{6.0in}{0}{80}{80}{-250}{-100}
\caption{}
\end{figure}

\clearpage

\begin{figure}
\plotfiddle{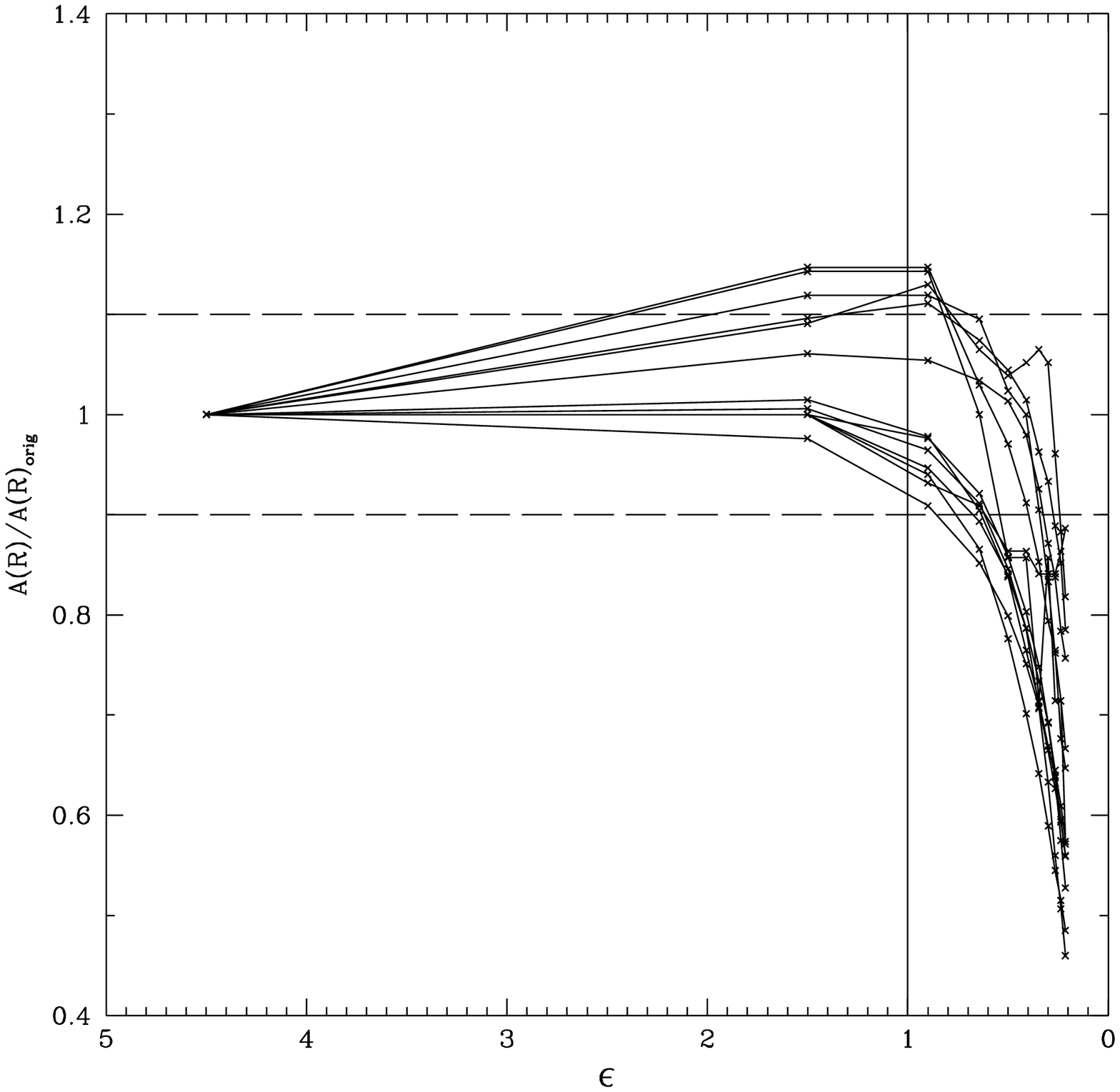}{6.0in}{0}{80}{80}{-250}{-100}
\caption{}
\end{figure}

\clearpage

\begin{figure}
\plotfiddle{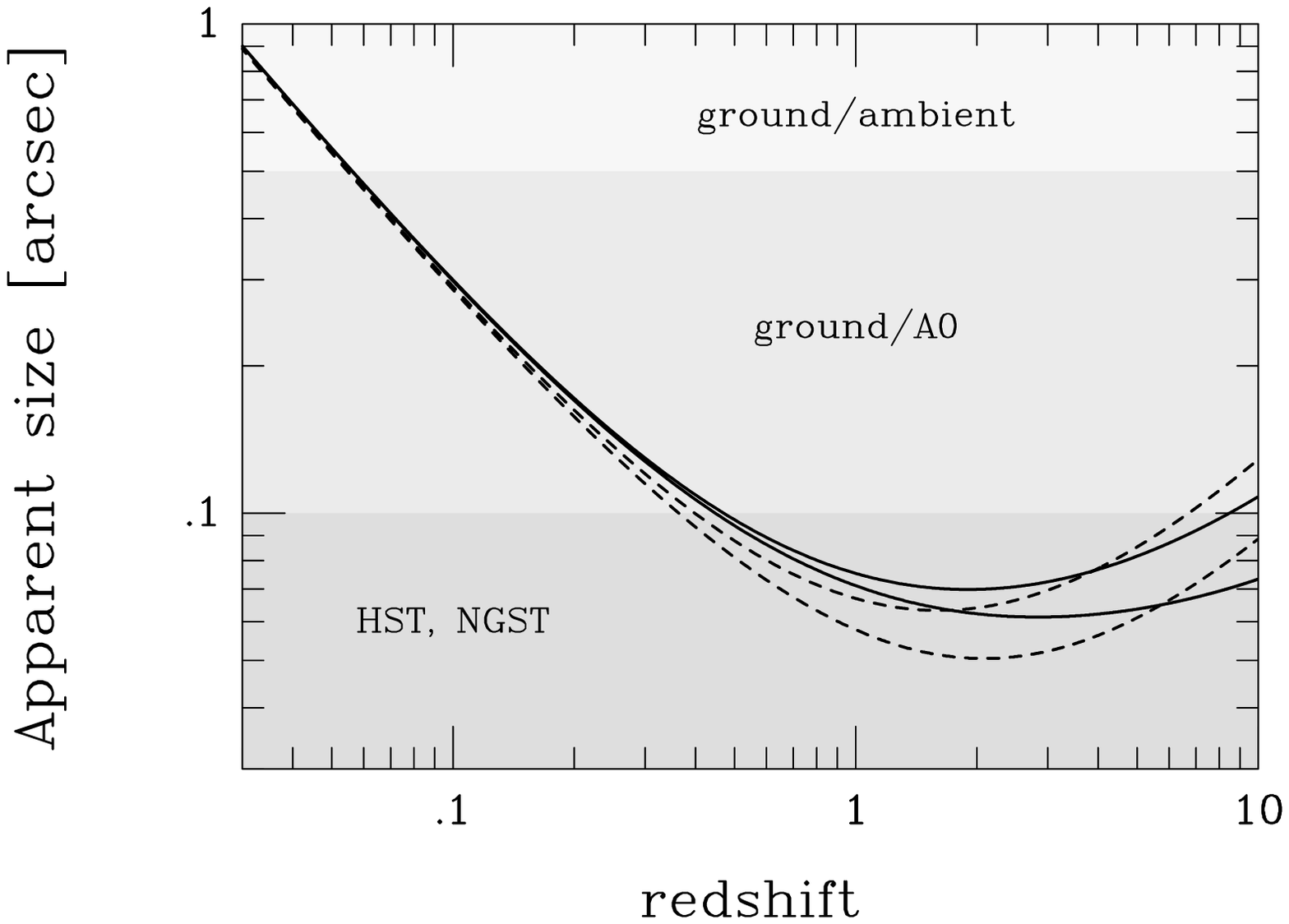}{6.0in}{0}{100}{100}{-300}{-100}
\caption{}
\end{figure}

\end{document}